\documentclass[acmsmall]{acmart}

\renewcommand\footnotetextcopyrightpermission[1]{} % removes footnote with conference information in first column
\setcopyright{acmcopyright}
\copyrightyear{2023}
\acmYear{2023}
\acmDOI{XXXXXXX.XXXXXXX}

\acmConference[DRAFT '23]{Make sure to enter the correct
  conference title from your rights confirmation emai}{June 03--05,
  2018}{Woodstock, NY}
\acmPrice{15.00}
\acmISBN{978-1-4503-XXXX-X/18/06}

\usepackage{xspace}
\usepackage{booktabs}   %% For formal tables:
\usepackage{subcaption} %% For complex figures with subfigures/subcaptions
\usepackage{fancyvrb}
\usepackage{xcolor}
\usepackage{wrapfig}
\usepackage{tikz}
\usepackage[clock]{ifsym}
\usepackage{array}

\usetikzlibrary{positioning}
\tikzset{main node/.style={circle,fill=gray!20,draw,minimum size=0.45cm,inner sep=0pt}, }
\tikzset{invisible/.style={circle,fill=white!20,draw,minimum size=0.0cm,inner sep=0pt}, }
\usepackage{multicol}
\usepackage{multirow}
\usepackage{listings}
\usepackage{syntax}
\usepackage{mathpartir}
\usepackage{mdframed}
\usepackage{multicol}
\usepackage[scaled=0.75]{FiraMono}

\lstdefinelanguage{plain}{
  basicstyle={\ttfamily},
}

\lstdefinelanguage{Slog}{
  basicstyle={\linespread{0.85}\ttfamily\small},       % the size of the fonts that are used for the code
  keywords = {},
  otherkeywords = {-->, [, ], !, ?, comp, <--, ...},
  morekeywords=[1]{-->},
  keywords = [2] {--},
  escapeinside={(@}{@)},          % if you want to add LaTeX within your code
  sensitive=true,
  numbers=left,
  morecomment=[l]{;},
  morecomment=[s]{\#|}{|\#},
  morestring=[b]",
  keywordstyle=\ttfamily\color[rgb]{0,.3,.7},
  commentstyle=\color[rgb]{0.133,0.545,0.133},
  stringstyle={\color[rgb]{0.75,0.49,0.07}},
  upquote=true,
  breaklines=true,
  breakatwhitespace=true,
  numbers=none,
  literate={lambda}{{$\lambda$}}1{langle}{{$\langle$}}1{rangle}{{$\rangle$}}1{dia}{{$\diamond$}}1{star}{{$\star$}}1{qmark}{{$?$}}1{col}{{$:$}}1{ell}{{$\ell$}}1{leftarrow}{{$\leftarrow$}}1
}

\lstdefinestyle{slogsty}{
    keywordstyle=\color{magenta},
    basicstyle=\scriptsize,
    breakatwhitespace=false,         
    breaklines=true,                 
    captionpos=b,                    
    keepspaces=true,                 
    numbers=left,                    
    numbersep=5pt,                  
    showspaces=false,                
    showstringspaces=false,
    showtabs=false,                  
    tabsize=2
}

\newcommand{\slog}[0]{\textsc{Slog}\xspace}
\newcommand{\lf}[0]{$\lambda\textit{LF}$\xspace}
\newcommand{\CC}{C\nolinebreak\hspace{-.05em}\raisebox{.4ex}{\tiny\bf +}\nolinebreak\hspace{-.10em}\raisebox{.4ex}{\tiny\bf +}}
\newcommand{\timeout}{{\tiny \StopWatchEnd{}}}

\newcommand{\core}{$\text{DL}_s$\xspace}

\definecolor{lightgray}{rgb}{0.3, 0.3, 0.3} 
\definecolor{lightblue}{rgb}{0.85, 0.95, 1.0} 
\definecolor{lightgreen}{rgb}{0.9, 1.00, 0.9} 
\definecolor{midorange}{rgb}{0.6, 0.40, 0.2}
\definecolor{mgreen}{rgb}{0.7, 0.90, 0.85}
\definecolor{black}{rgb}{0.0, 0.0, 0.0}
\newcommand{\rtag}[1]{\textcolor{lightgray}{\bf{\texttt{#1}}}}
\newcommand{\huhclause}[1]{{\setlength\fboxsep{0pt}\colorbox{lightblue}{#1}}}
\newcommand{\bangclause}[1]{{\setlength\fboxsep{0pt}\colorbox{lightgreen}{#1}}}
\newcommand{\comm}[1]{\textcolor{midorange}{#1}}

\begin{document}

%%
%% The "author" command and its associated commands are used to define
%% the authors and their affiliations.
%% Of note is the shared affiliation of the first two authors, and the
%% "authornote" and "authornotemark" commands
%% used to denote shared contribution to the research.

%% ACMART NOT IEEE:::::
\author{Thomas Gilray}
\email{gilray@uab.edu}
\affiliation{%
  \institution{University of Alabama at Birmingham}
  \country{USA}
}

\author{Arash Sahebolamri}
\email{asahebol@syr.edu}
\affiliation{%
  \institution{Syracuse University}
  \country{USA}
}

\author{Sidharth Kumar}
\email{sid14@uab.edu}
\affiliation{%
  \institution{University of Alabama at Birmingham}
  \country{USA}
}

\author{Kristopher Micinski}
\email{kkmicins@syr.edu}
\affiliation{%
  \institution{Syracuse University}
  \country{USA}
}

%%
%% The "title" command has an optional parameter,
%% allowing the author to define a "short title" to be used in page headers.
\title{Higher-Order, Data-Parallel Structured Deduction}

%%
%% By default, the full list of authors will be used in the page
%% headers. Often, this list is too long, and will overlap
%% other information printed in the page headers. This command allows
%% the author to define a more concise list
%% of authors' names for this purpose.
\renewcommand{\shortauthors}{Gilray et al.}

%%
%% The abstract is a short summary of the work to be presented in the
%% article.
\begin{abstract}
  State-of-the-art Datalog engines include expressive features such as
  ADTs (structured heap values), stratified aggregation and negation,
  various primitive operations, and the opportunity for further extension
  using FFIs. Current parallelization approaches for state-of-art Datalogs
  target shared-memory locking data-structures using conventional
  multi-threading, or use the map-reduce model for distributed computing.
  Furthermore, current state-of-art approaches cannot scale to formal systems
  which pervasively manipulate structured data due to their lack of
  indexing for structured data stored in the heap.

  In this paper, we describe a new approach to data-parallel structured
  deduction that involves a key semantic extension of Datalog to permit
  first-class facts and higher-order relations via defunctionalization,
  an implementation approach that enables parallelism uniformly both
  across sets of disjoint facts and over individual facts with nested structure.
  We detail a core language, \core{}, whose key invariant (subfact closure)
  ensures that each subfact is materialized as a top-class fact. We
  extend \core{} to \slog{}, a fully-featured language whose forms
  facilitate leveraging subfact closure to rapidly implement
  expressive, high-performance formal systems. We demonstrate
  \slog{} by building a family of control-flow analyses from abstract machines, systematically,
  along with several implementations of classical type systems (such as STLC and LF).
  We performed experiments on EC2, Azure, and ALCF's Theta at up to 1000 threads,
  showing orders-of-magnitude scalability improvements versus competing state-of-art systems.
\end{abstract}

%%
%% The code below is generated by the tool at http://dl.acm.org/ccs.cfm.
%% Please copy and paste the code instead of the example below.
%%

%%
%% Keywords. The author(s) should pick words that accurately describe
%% the work being presented. Separate the keywords with commas.
%% A "teaser" image appears between the author and affiliation
%% information and the body of the document, and typically spans the
%% page.

%%
%% This command processes the author and affiliation and title
%% information and builds the first part of the formatted document.
\maketitle

\section{Structured Declarative Reasoning}
\label{sec:intro}
Effective programming languages permit their user to write high-performance code in a manner that is as close to the shape of her own thinking as possible. 
A long-standing dream of our field has been to develop especially high-level \emph{declarative} languages that help bridge this gap between specification and implementation. Declarative programming permits a user to provide a set of high-level rules and declarations that offer the sought-after solution as a latent implication to be materialized automatically by the computer. 
The semantics of a declarative language does the heavy lifting in operationalizing this specification for a target computational substrate---one with its own low-level constraints and biases. Modern computers provide many threads of parallel computation, may be networked to further increase available parallelism, and are increasingly virtualized within ``cloud'' services. To enable scalable cloud-based reasoning, the future of high-performance declarative languages must refine their suitability on both sides of this gulf: becoming both more tailored to human-level reasoning and to modern, massively-parallel, multi-node machines.

Logic-programming languages that extend Datalog have seen repeated resurgences in interest since their inception, each coinciding with new advances in their design and implementation.
For example, Bddbddb~\cite{whaley2005using} suggested that binary decision diagrams (BDDs) could be used to compress relational data while permitting fast algebraic operations such as relational join, but required \emph{a priori} knowledge of efficient BDD-variable orderings to enable its compression, which proved to be a significant constraint.
LogicBlox and Souffl\'e~\cite{antoniadis2017porting,10.1007/978-3-319-41540-6} have since turned research attention back to semi-na\"ive evaluation over extensional representations of relations, using compression techniques sparingly (i.e., compressed prefix trees) and focusing on the development of high-performance shared-memory data structures.
Souffl\'e represents the current state of the art at a low thread count, but struggles to scale well due to internal locking and its coarse-grained approach to parallelism.
RadLog (i.e., BigDatalog)~\cite{bigdl} has proposed scaling deduction to many-thread machines and clusters using Hadoop and the map-reduce paradigm for distributed programming. Unfortunately, map-reduce algorithms suffer from a (hierarchical) many-to-one collective communication bottleneck and are increasingly understood to be insufficient for leading high-performance parallel-computing environments~\cite{Anderson:2017,SparkBadMPIGood}.

Most modern Datalogs are Turing-equivalent extensions, not simply finite-domain first-order HornSAT, offering stratified negation, abstract data-types (ADTs), ad hoc polymorphism, aggregation, and various operations on primitive values.
Oracle's Souffl\'e has added flexible pattern matching for ADTs, and Formulog~\cite{formulog-bembenek2020} shows how these capabilities can be used to perform deductive inference of formulas; it seems likely future (extended) Datalogs will be used to implement symbolic execution and formal verification in a scalable, parallel manner.    

In this paper, we introduce a new approach to simultaneously improve the expressiveness and data-parallelism of such deductive logic-programming languages. Our approach has three main parts: (1) a key semantic extension to Datalog, \emph{subfacts} and a \emph{subfact-closure property}, that is (2) implemented uniformly via ubiquitous fact interning, supported within relational algebra operations that are (3) designed from the ground-up to automatically balance their workload across available threads, using MPI to address the available data-parallelism directly. We show how our extension to Datalog permits deduction of structured facts, defunctionalization and higher-order relations, and more direct implementations of abstract machines (CEK, Krivine's, CESK), rich program analyses ($k$-CFA, $m$-CFA), and type systems. We detail our implementation approach and evaluate it against the best current Datalog systems, showing improved scalability and performance.

We offer the following contributions to the literature:
\begin{enumerate}
\item An architecture for extending Datalog to structured recursive data and higher-order relations, uniform with respect to parallelism, allowing inference of tree-shaped facts which are indexed and data-parallel both horizontally (across facts) and vertically (over subfacts).
\item A formalism of our core language, relationship to Datalog, and equivalence of its model theoretic and fixed-point semantics, mechanized in Isabelle/HOL.
\item A high-performance implementation of our system, \slog{}, with a compiler, REPL, and runtime written in Racket (10.6kloc), Python (2.5kloc), and \CC{} (8.5kloc).
\item An exploration of \slog{}'s applications in the engineering of formal systems, including program analyses and type systems. We include a
%an end-to-end
  presentation of the systematic development of program analyses from corresponding abstract-machine interpreters---the abstracting abstract machines (AAM) methodology---where
  %showing
  each intermediate step in the AAM process may also be written using \slog{}.
\item An evaluation comparing \slog{}'s performance against Souffl\'e and RadLog on EC2 and Azure, along with a strong-scaling study on the ALCF's Theta supercomputer which shows promising strong scaling up to 800 threads. We observe improved scaling efficiency and performance at-scale, compared with both Souffl\'e and RadLog, and better single-thread performance vs. Souffl\'e when comparing \slog{} subfacts to Souffl\'e ADTs.
\end{enumerate}

\section{Slog: Declarative Parallel Deduction of Structured Data}
\label{sec:over}
For Datalogs used in program analysis, manipulation of abstract syntax trees (ASTs) is among the most routine tasks.
Normally, to provide such ASTs as input to a modern Datalog engine, one first requires an external flattening tool that walks the richly structured syntax tree and produces a stream of flat, first-order facts to be provided as an input database. For example, the Datalog-based Java-analysis framework DOOP~\cite{Bravenboer:2009, 10.1145/1639949.1640108}, ported to Souffl\'e~\cite{antoniadis2017porting} in 2017, has a substantial preprocessor (written in Java) to be run on a target JAR to produce an input database of AST facts for analysis.

A key observation that initially motivated our work into this subject was that although this preparatory
transformation is required to provide an AST as a database of first-order facts,
the same work could not be done from within these Datalogs because it required
generating unique identities (i.e., pointers to intern values) for inductively defined terms. In fact, any work generating ASTs as facts can not be
done within Datalog itself but must be an extension to the language. Consider the pair of nested expressions
that form an identity function:

%\vspace{-0.65cm}
\begin{multicols}{3}
  \colorbox{white}{\texttt{(\rtag{lam} "x" (\rtag{ref} "x"))}}

  \hspace{0.5cm} $\overset{\text{flattens}}{\longrightarrow}$ %\hspace{-0.5cm}
  
  \begin{minipage}{\linewidth}
  \texttt{(= lam-id (\rtag{lam} "x" ref-id))}
  
  \texttt{(= ref-id (\rtag{ref} "x"))}
  \end{minipage}
\end{multicols}

Supplying unique intern values \texttt{lam-id} and \texttt{ref-id} as an extra column for those relations, and thus permitting them to be linked together, is the substance of this preparatory transformation. Our language, \slog{}, proposes this interning behavior for facts be ubiquitous, accounted for at every iteration of relational algebra used to implement the underlying HornSAT fixed point. 

In Souffl\'e, the language has more recently provided abstract data-type (ADT) declarations and struct/record types for heap-allocated values which can be built up into ASTs
or other such structured data. These datatypes must be declared and can then be used as \$ expressions within rules; e.g., \texttt{\$lam("x",\$ref("x"))}.
The downside of these ADTs in Souffl\'e is that they are not treated as facts for the purposes of triggering rules and are
not indexed as facts, which would permit more efficient access patterns.

Our language, \slog{}, respects a \emph{subfact closure property}: every subfact is itself a first-class fact in the language and every top-level fact (and subfact) is a first-class value and has a unique identity (as an automatic column-$0$ value added to the relation). A clause \texttt{(\rtag{foo} x y)} in \slog{}, always has an implied identity column and is interpreted the same as \texttt{(= \_ (\rtag{foo} x y))} if it's missing (where underscore is a wildcard variable). A nested pair of linked facts like \texttt{(\rtag{foo} x (\rtag{bar} y) z)} is desugared as \texttt{(= \_ (\rtag{foo} x id z))} and \texttt{(= id (\rtag{bar} y))}. Thus we can represent an identity function's AST in \slog{} as the directly nested fact and subfact \texttt{(\rtag{lam} "x" (\rtag{ref} "x"))}; under the hood this will be equivalent to two flat facts with a 0-column id provided by an interning process that occurs at the discovery of each new \slog{} fact.

\begin{itemize}
\item In \slog{}, each structurally unique fact/subfact has a unique intern-id stored in its 0 column so it may be referenced as another fact's subfact and treated as a first-class value.
\item In \slog{}, all data is at once a first-class fact (able to trigger rule evaluation), a first-class value (able to be referenced by other facts/values), and a first-class thread of execution (treated uniformly by a data-parallel MPI backend that dynamically distributes the workload spatially within, and temporally across, fixed-point iterations).
\end{itemize}

With subfacts as first-class citizens of the language (see Section~\ref{sec:semantics} for details), able to trigger rules, various useful idioms emerge in which a subfact triggers a response from another rule via an enclosing fact (see Section~\ref{sec:semantics:extensions} for extensions and idioms). Using these straightforward syntactic extensions enables a wide range of deduction and reasoning systems (see Section~\ref{sec:apps} for a discussion of applications in program analyses and type systems). Because subfacts are first-class in \slog{}, rules that use them will naturally force the compiler to include appropriate indices enabling efficient access patterns, and represent thread joins in the natural data-parallelism \slog{} exposes. As a result, we are able to show a deep algorithmic improvement and parallelism over current state-of-art systems in the implementations of analyses we generate (see Section~\ref{sec:eval} for our evaluation with apples-to-apples comparisons against Souffl\'e and RadLog). In some experiments, \slog{} finishes in 4--8 seconds with Souffl\'e taking 1--3 hours---attesting to the importance of subfact indices. In others, we observe efficient strong-scaling on up to hundreds of threads, showing the value of our data-parallel backend.

\section{Structurally Recursive Datalog}
\label{sec:semantics}
%\lstset{language=Slog}
%
The core semantic difference between \slog{} and Datalog is to allow
structurally recursive, first-class facts.  This relatively minor
semantic change enables both enhanced expressivity (naturally
supporting a wide range of Turing-equivalent idioms, as we demonstrate
in Section~\ref{sec:apps}) and anticipates compilation to parallel
relational algebra (which interns all facts and distributes facts via
their intern key). In this section, we present the formal semantics of
a language we call Structurally Recursive Datalog (henceforth
\core{}), the core language extending Datalog to which \slog{}
programs compile. All of the definitions related to \core{} have been
formalized, and all of the lemmas and theorems presented in this
section have been formally proven in Isabelle.

\paragraph*{Syntax}
\begin{wrapfigure}{r}{6cm}
\vspace{-0.5cm}
\begin{grammar}
  <Prog>   ::= <Rule>*
  
  <Rule>   ::= <Clause> $\leftarrow$ <Clause>*
  
  <Clause> ::= (tag <Subcl>*)
  
  <Subcl>  ::= (tag <Subcl>*) | <Var> | <Lit>
  
  <Lit>    ::= <Number> | <String> | ...
\end{grammar}
\caption{Syntax of \core{}: \textit{tag} is a relation name.}
\label{fig:dls-syntax}
\end{wrapfigure}
The syntax of \core{} is shown in Figure~\ref{fig:dls-syntax}. As in
Datalog, a \core{} program is a collection of Horn clauses. Each
rule $R$ contains a set of body clauses and a head clause, denoted by
$\textit{Body}(R)$ and $\textit{Head}(R)$ respectively. \core{} (and
\slog{}) programs must also be well-scoped: variables appearing in a
head clause must also be contained in the body.
  
We define a strict syntactic subset of \core{}, \textit{DL} as the
restriction of \core{} to clauses whose arguments are literals (i.e.,
${\langle\textit{Subcl}\rangle}_{\textit{DL}} ::= \langle\textit{Var}\rangle \mid
\langle\textit{Lit}\rangle$). This subset (and its semantics) corresponds to
Datalog.

\paragraph*{Fixed-Point Semantics}

The fixed-point semantics of a \core{} program $P$ is given via the
least fixed point of an \emph{immediate consequence} operator
$\textit{IC}_P : \textit{DB} \rightarrow \textit{DB}$. Intuitively,
this immediate consequence operator derives all of the immediate
implications of the set of rules in $P$. A
database $\textit{db}$ is a set of facts ($\textit{db} \in \textit{DB}
= \mathcal{P}(\textit{Fact})$). A fact is a clause without variables:
\begin{align*}
\textit{Fact} &::= (\textit{tag}\ \textit{Val}^*) & \quad
\textit{Val}  &::= (\textit{tag}\ \textit{Val}^*)\ \mid \textit{Lit}
\end{align*}

In Datalog, $\textit{Val}$s are restricted to a finite set of atoms
($\textit{Val}_{\textit{DL}} ::= \textit{Lit}$). To define $IC_P$, we
first define the immediate consequence of a rule $IC_R : DB
\rightarrow DB$, which supplements the provided database with all the
facts that can be derived directly from the rule given the available
facts in the database:

\begin{align*}    
\textit{IC}_R(db) \triangleq\ & db\ \cup \bigcup \bigl\{ \textit{unroll}(\textit{Head}(R)\big[\overrightarrow{v_i\backslash x_i}\big]) | \\
& \{\overrightarrow{x_i \rightarrow v_i}\} \subseteq (\textit{Var} \times \textit{Val})\ \wedge
\textit{Body}(R)\big[\overrightarrow{v_i \backslash x_i}\big] \subseteq db \bigr\}
\end{align*}

The $unroll$ function has the following definition:

\begin{centering}
  {\centering$\begin{array}{rcl}
\quad\quad\quad \textit{unroll}\bigl((\textit{tag}\ \textit{item}_1\ ...\ \textit{item}_n)\bigr) &\triangleq&
\{(\textit{tag}\ \textit{item}_1\ ...\ \textit{item}_n)\} \cup \bigcup_{i \in {1 ... n}} \textit{unroll}(\textit{item}_i)   \\
\textit{unroll}(v)_{v \in \textit{Lit}} &\triangleq& \{\}
    \end{array}$}
\end{centering}

The purpose of the unroll function is to ensure that all nested facts are included in the database as well, a property we call \emph{subfact-closure}. This property is crucial to the semantics of \core{} (and \slog{}), because in \core{}, each nested fact is a fact in own right, and not merely a carrier of structured data. Later sections (starting in section~\ref{sec:semantics:extensions}) illustrate the importance of subfact closure by demonstrating how  we utilize this behavior to construct idioms that make programming in \slog{} more convenient.

The immediate consequence of a program is the union of the immediate
consequence of each of its constituent rules, $
\textit{IC}_P(\textit{db}) \triangleq \textit{db} \cup \bigcup_{R \in
  P} \textit{IC}_R(\textit{db})$. Observe that $\textit{IC}_P$ is
monotonic over the the lattice of databases whose bottom element is
the empty database. Therefore, if $\textit{IC}_P$ has any fixed
points, it also has a least fixed point \cite{tarski1955lattice}.
Iterating to this least fixed point directly gives us a na\"ive, incomputable
fixed-point semantics for \core{} programs.
Unlike pure Datalog, existence of a finite fixed point is not guaranteed in
\core{}. This is indeed a reflection of the fact that \core{} is
Turing-complete. The \core{} programs whose immediate consequence
operators have no finite fixed points are non-terminating.

As discussed earlier, all \slog{} databases must be subfact-closed (i.e. all subfacts are first-class facts). We can show that the least fixed point of the immediate consequence operator has the property that it is subfact-closed.
\begin{lemma}
(Formalized in Isabelle.) The least fixed point of $\textit{IC}_P$ is subfact-closed. 
\end{lemma}

It is worth pointing out that the fixed point semantics of Datalog is similar, the only difference being that the $\textit{unroll}$ function is not required, as Datalog clauses do not contain subclauses.

\paragraph*{Model Theoretic Semantics}

The model theoretic semantics of \core{} closely follows the model
theoretic semantics of Datalog, as presented in, e.g.,
\cite{ceri1989you-datalog}. The \emph{Herbrand universe} of a \core{}
program is the set of all of the facts that can be constructed from
the relation symbols appearing in the program. Because \core{} facts
can be nested, the Herbrand universe of any nontrivial \core{} program
is infinite. One could for example represent natural numbers in
\core{} using the zero-arity relation \lstinline{Zero} and the unary
relation \lstinline{Succ}. The Herbrand universe produced by just
these two relations, one zero arity and one unary, is inductively infinite.

A \emph{Herbrand Interpretation} of a \core{} program is any subset of its Herbrand universe that is subfact-closed. In other words, if $I$ is a Herbrand Interpretation, then $I = \bigcup \{\textit{unroll}(f) \linebreak |\ f \in I\}$. For Datalog, the Herbrand Interpretation is defined similarly, with the difference that subfact-closure is not a requirement for Datalog, as Datalog facts do not contain subfacts.

Given a Herbrand Interpretation $I$ of a \core{} program $P$, and a rule $R$ in $P$, we say that $R$ is true in $I$ ($I \models R$) iff for every substitution of variables in $R$ with facts in $I$, if all the body clauses with those substitutions are in $I$, so is the head clause of $R$ with the same substitutions of variables.
\begin{align*}  
& I \models R\ ~\textrm{iff} ~
 \forall \{\overrightarrow{x_i \rightarrow v_i}\}\ .\ \textit{Body}(R)\big[\overrightarrow{v_i\backslash x_i}\big] \subseteq I \longrightarrow \textit{Head}(R)\big[\overrightarrow{v_i\backslash x_i}\big] \in I 
\end{align*}

If every rule in $P$ is true in $I$, then $I$ is a \emph{Herbrand model} for $P$. The denotation of $P$ is the intersection of all Herbrand models of $P$. We define $\mathbf{M}(P)$ to be the set of all Herbrand models of $P$, and $D(P)$ to be the denotation of $P$. We then have $D(P) \triangleq \!\!\!\!\!\!\bigcap\limits_{I\in\ \mathbf{M}(P)}\!\!\!\!\! I$. It can be shown that such an intersection is a Herbrand model itself:
\begin{lemma}
The intersection of a set of Herbrand models is also a Herbrand model.
\end{lemma}

Unlike Datalog, nontrivial \core{} programs have Herbrand universes
that are infinite. Thus, a \core{} program may have only infinite
Herbrand models. If a \core{} program has no finite Herbrand models,
its denotation is infinite and so no fixed-point may be finitely
calculated using the fixed-point semantics. We now relate the operational
semantics of \core{} to its model-theoretic semantics.

\paragraph*{Equivalence of Model-Theoretic and Fixed-Point Semantics}

To show that the model-theoretic and fixed-point semantics of \core{}
compute the same Herbrand model, we need to show that the least fixed
point of the immediate consequence operator is equal to the
intersection of all the Herbrand models for any program. We start by
proving the following lemmas (proved in Isabelle; proofs elided for
space).

\begin{lemma}
Herbrand models of a \core{} program are fixed points of the immediate consequence operator.    
\end{lemma}

\begin{lemma}
Fixed points of the immediate consequence operator of a \core{} program that are subfact-closed are Herbrand models of the program.    
\end{lemma}

\begin{wrapfigure}{l}{6.75cm}
%\begin{tabular}{ c c }

\setlength{\grammarindent}{7em} % increase separation between LHS/RHS 
% \begin{minipage}{.50\textwidth}
\begin{grammar}
<toplvl-rule> ::= <rule> | <hclause> 

<rule> ::= "[" <hd-item>* "<--" <bd-item>* "]"
      \alt "[" <bd-item>* "-->" <hd-item>* "]" 

<bd-item> ::= <rule> | <bclause> 

<hd-item> ::= <rule> | <hclause> 

<bclause> ::= "("<tag> <ibclause>*")"  
         \alt "(=" <var> "("<tag> <ibclause>*"))"

<hclause> ::= "("<tag> <ihclause>*")" 
         \alt "(=" <var> "("<tag> <ihclause>*"))"

<atom> ::= <var> | <lit>

<lit> ::= <string> | <number>

% \end{grammar}
% \end{minipage}
% \begin{minipage}{.45\textwidth}
% \begin{grammar}
<ihclause> ::= "("<tag> <ihclause>*")" 
          \alt "?("<tag> <ibclause>*")"
          \alt "{"<tag> <ibclause>*"}" 
          \alt "["<hlist-item>*"]" 
          \alt "?["<blist-item>*"]"
          \alt <atom>

<ibclause> ::= "("<tag> <ibclause>*")" 
          \alt "!("<tag> <ihclause>*")" 
          \alt "{"<tag> <ibclause>*"}" 
          \alt "["<blist-item>*"]" 
          \alt "!["<hlist-item>*"]"
          \alt <atom>

% TODO remove list-related grammar if lists are taken out of the prose.
<hlist-item> ::= <ihclause> | <ihclause> "…"

<blist-item> ::= <ibclause> | <ibclause> "…"
\end{grammar}
% \end{minipage}

%&
%\\
%\end{tabular}
\caption{The syntax of \slog{}. \synt{var} is the set of variables, and \synt{tag} is the set of relation names. A few syntactic forms, including disjunction, have been elided.}
\label{fig:syntax}
\vspace{-1.5cm}
\end{wrapfigure}

By proving that the Herbrand models and subfact-closed fixed points of
the immediate consequence operator are the same, we conclude that the
least fixed point of the immediate consequence operator
$\textit{IC}_P$ (a subfact-closed database) is equal to the
intersection of all its Herbrand models.

\begin{theorem}
The model theoretic semantics and fixed point semantics of \core{} are equivalent.
\end{theorem}

Proof sketch: Form the lemma that all Herbrand models are fixed points of $\textit{IC}_P$, we conclude that $D(P)$ is a superset of the intersection of all the fixed points. We know that the least fixed point of $\textit{IC}_P$ (which we'll call $\textit{LFP}_P$) is a subset of the intersection of all the fixed points. We therefore have $\textit{LFP}_P \subseteq D(P)$. From the fact the $\textit{LFP}_P$ is a Herbrand model, we conclude that $D(P) \subseteq \textit{LFP}_P$. Putting these facts together, we conclude that $\textit{LFP}_P = D(P)$.
% From these lemmas, we know that Herbrand models and fixed points of the immediate consequence operator are the same. We therefore need to show that the intersection of all the fixed points of the immediate consequence operator is the least fixed point of the operator. From the definition of the least fixed point, it immediately follows that if $db_1$ and $db_2$ are fixed points of the operator, $IC_P(db_1 \cap db_2) \subseteq db_1 \cap db_2$. Conversely, from the definition of $IC_P$, it follows that for all $db$, $db \subseteq IC_P(db)$. Putting these two facts together, we have: for all fixed points $db_1$ and $db_2$ of $IC_P$, $IC_P(db_1 \cap db_2) = db_1 \cap db_2$.

\renewcommand{\multicolsep}{3pt plus 1pt minus 1pt}

% https://tex.stackexchange.com/questions/24886/which-package-can-be-used-to-write-bnf-grammars
% https://mirrors.rit.edu/CTAN/macros/latex/contrib/mdwtools/syntax.pdf
\renewcommand{\ulitleft}{\bf\ttfamily\frenchspacing}
\renewcommand{\ulitright}{}

\subsection{Key extensions to the core language}
\label{sec:semantics:extensions}
With subfacts, a common idiom becomes for a subfact to appear in the body of a rule, while its surrounding fact and any associated values are meant to appear in the head. For these cases, we use a ? clause, an s-expression marked with a ``?'' at the front to indicate that although it may appear to be a head clause, it is actually a body clause and the rule does not fire without this fact present to trigger it. The following rule says that if a \texttt{(\rtag{ref} x)} AST exists, then x is a free variable with respect to it.
\begin{Verbatim}[baselinestretch=0.75,commandchars=\\\{\}]
(\rtag{free} \huhclause{?(\rtag{ref} x)} x)
\end{Verbatim}
which desugars to the rule
\begin{Verbatim}[baselinestretch=0.75,commandchars=\\\{\}]
[(= e-id (\rtag{ref} x)) --> (\rtag{free} e-id x)]
\end{Verbatim}
exposing that the ? clause is an implicit body clause. But if there are
no body clauses apart from the ? clauses, the rule may be written without
square braces and an arrow to show direction.

\begin{wrapfigure}{r}{6cm}
\vspace{-0.25cm}
\begin{Verbatim}[baselinestretch=0.75,commandchars=\\\{\}]
[(=/= x y) (\rtag{free} Eb y)
 --> (\rtag{free} \huhclause{?(\rtag{lam} x Eb)} y)]
[(or (\rtag{free} Ef x) (\rtag{free} Ea x))
 --> (\rtag{free} \huhclause{?(\rtag{app} Ef Ea)} x)
\end{Verbatim}
\end{wrapfigure}
Two more rules are needed to define a free-variables analysis.
The second of these shows another extension: disjunction in the body of
a rule is pulled to the top level and splits the rule into multiple rules.
In this case, there is both a rule saying that a free variable in \texttt{Ef}
is free in the application and a rule saying that a free variable in \texttt{Ea}
is free in the application.

Another core mechanism in \slog{} is to put head clauses in position where a body clause is expected.
Especially because an inner clause can be \emph{responded to} by a fact surrounding it, or by rules producing that fact,
being able to emit a fact on-the-way to computing a larger rule is what permits natural-deduction-style rules through a kind of rule
splitting, closely related to continuation-passing-style (CPS) conversion~\cite{appel2007compiling}.
A ! clause, under a ? clause or otherwise in the position of a body clause,
is a clause that will be deduced as the surrounding rule is evaluated, so long as any ? clauses are satisfied and any subexpressions
are ground (any clauses it depends on have been matched already). These ! clauses are intermediate head clauses; technically the head clauses of subrules, which they are compiled into internally. 

Consider the example in Figure~\ref{fig:natural-deduction-plus}, which lets us prove an arithmetic statement like\newline\texttt{(plus (plus (nat 1) (nat 2)) (nat 1)) $\Downarrow$ 4}.
We can construe this rule in a few ways, as written. It could be that both the expression and value should be provided and are proved according to these rules, or it could be treated as a calculator, with the expression provided as input.

\vspace{-0.4cm}
\begin{figure*}[h]
\begin{multicols}{2}
\begin{Verbatim}[baselinestretch=0.75,commandchars=\\\{\}]
(\rtag{interp} \huhclause{?(\rtag{do-interp} (\rtag{nat} n))} n)

  
[(\rtag{interp} \bangclause{!(\rtag{do-interp} e0)} v0)
 (\rtag{interp} \bangclause{!(\rtag{do-interp} e1)} v1)
 (+ v0 v1 v) 
 --> \comm{;-------- [plus]}
 (\rtag{interp} \huhclause{?(\rtag{do-interp} (\rtag{plus} e0 e1))} v)]
\end{Verbatim}
\columnbreak
\[
  \frac{\ }{\texttt{(nat $n$)} \Downarrow n}\text{\small[nat]}
\]
  \vspace{0.75cm}
\[
 \frac{e_0 \Downarrow v_0 \hspace{1cm} e_1 \Downarrow v_1 \hspace{1cm} v=v_0+v_1}{(\texttt{plus}\ e_0\ e_1) \Downarrow v} \text{\small[plus]}
\]
\end{multicols}
\caption{Natural-deduction-style reasoning with ! clauses in \slog{}.}
\label{fig:natural-deduction-plus}
\end{figure*}
\vspace{-0.4cm}

\begin{wrapfigure}{r}{5.75cm}
\vspace{-0.3cm}
\begin{Verbatim}[baselinestretch=0.75,commandchars=\\\{\}]

[(\rtag{interp} \bangclause{!(\rtag{do-interp} e0)} v0)
 (\rtag{interp} \bangclause{!(\rtag{do-interp} e1)} v1) 
 --> \comm{;-------- [plus]}
 (\rtag{interp} \huhclause{?(\rtag{do-interp} (\rtag{plus} e0 e1))}
         \{+ v0 v1\})]




(\rtag{append} \huhclause{?(\rtag{do-append} [] ls)} ls)

[(\rtag{append} \bangclause{!(\rtag{do-append} ls0 ls1)} ls')
 -->
 (\rtag{append} \huhclause{?(\rtag{do-append} [x ls0 ...] ls1)}
         [x ls' ...])]

\comm{; or ind. case could even be written:}
(\rtag{append} \huhclause{?(\rtag{do-append} [x ls0 ...] ls1)}
        [x
         \{\rtag{append} \bangclause{!(\rtag{do-append} ls0 ls1)}\}
         ...])
\end{Verbatim}
\end{wrapfigure}
Subclauses, written with parentheses, are treated as top-level clauses whose id column value is unified at the position of the subclause. Another common use for a relation is as a function, or with a designated output column, deterministic or not, so \slog{} also supports this type of access via \{ \!\} inner clauses, which have their final-column value unified at the position of the curly-brace subclause. For example, the rule in Figure~\ref{fig:natural-deduction-plus} could also have been written as below, with the clause \texttt{\{+ v0 v1\}} in place of variable \texttt{v}.
This example illustrates that this syntax can also be used for built-in relations like \texttt{+}.

Putting this all together and adding \slog{}'s built-in list syntax---currently implemented as linked-lists of \slog{} facts in the natural way---we can implement rules for appending lists, a naturally direct-recursive task due to a linked list naturally having its first element at its front, so a second list can only be appended to the back of the first list, and the front element onto the front of that.

\section{Applications}
\label{sec:apps}
In this section, we will examine several related applications of \slog{}: implementing reduction systems, natural deduction systems, AAM-based program analyses, and natural-deduction-style type systems. 

We start with a $\lambda$-calculus interpreter. Let's observe how $\beta$-reduction can be defined via capture-avoiding substitution. If a \textbf{do-subst} fact is emitted where a reference to variable \texttt{x} is being substituted with expression \texttt{E}, associate it in the \textbf{subst} relation with \texttt{E}:
\begin{Verbatim}[baselinestretch=0.8,commandchars=\\\{\}]
(\rtag{subst} \huhclause{?(\rtag{do-subst} (\rtag{ref} x) x E)} E)
\end{Verbatim}
However, if \texttt{x} and \texttt{y} are distinct variables, the substitution yields expression \texttt{(ref x)} unchanged:
\begin{Verbatim}[baselinestretch=0.8,commandchars=\\\{\}]
[(=/= x y) --> (\rtag{subst} \huhclause{?(\rtag{do-subst} (\rtag{ref} x) y E)} (\rtag{ref} x))]
\end{Verbatim}
Recall that ?-clauses are body clauses, so these rules could also have been written more verbosely:
\begin{Verbatim}[baselinestretch=0.8,commandchars=\\\{\}]
[(= d (\rtag{do-subst} (\rtag{ref} x) x E)) --> (\rtag{subst} d E)]
[(=/= x y) (= d (\rtag{do-subst} (\rtag{ref} x) y E)) --> (\rtag{subst} d (\rtag{ref} x))]
\end{Verbatim}

At a lambda, where the formal parameter shadows the variable being substituted, its scope ends and substitution stops:
\begin{Verbatim}[baselinestretch=0.8,commandchars=\\\{\}]
(\rtag{subst} \huhclause{?(\rtag{do-subst} (\rtag{lam} x Ebody) x E)} (\rtag{lam} x Ebody))
\end{Verbatim}
If the variable does not match and is not free in the (argument) expression \texttt{E}, substitution may continue under the lambda, triggered by a ! clause:
\begin{Verbatim}[baselinestretch=0.8,commandchars=\\\{\}]
[(=/= x y)  ~(free E x)   
 --> (\rtag{subst} \huhclause{?(\rtag{do-subst} (\rtag{lam} x Ebody) y E)}
            (\rtag{lam} x \{\rtag{subst} \bangclause{!(\rtag{do-subst} Ebody y E)}\}))]
\end{Verbatim}
Three further syntactic extensions are being used in this rule. First off, the negated \texttt{\textasciitilde{}(free E x)} clause in the body requires that the compiler stratify computation of \texttt{free}, as normal when adding otherwise nonmonotonic rule dependance to Datalog programs.

Second, the process of rewriting the lambda body is triggered by the establishment of a \textbf{\texttt{do-subst}} fact via a ! clause. These ! clauses generate facts on-the-fly during rule evaluation, allowing other rules to hook-in by generating a fact to trigger them (using a ! clause) and expecting a response in the surrounding body clauses. These ! clauses are implemented by generating a subrule whose head clause is the intermediate ! clause and whose body contains all body clauses the ! clause depends upon, along with and any ? clauses in the rule (which are always required to trigger any rule). In this case, term-substitution rules respond to the \rtag{do-subst} request via the \textbf{\texttt{subst}} relation, as queried here by the \{ \!\} expression in \texttt{\{\rtag{subst} \bangclause{!(\rtag{do-subst} Ebody y E})}\}.

Third, this \{ \!\} syntax allows for looking up the final column of a relation by providing all but the final-column value. \texttt{(foo x \{bar y\})} desugars into \texttt{(and (foo x z) (bar y z))}, allowing for the looked-up value to be unified with the position of the \{ \!\} expression in a natural way. Do note that the relation need not actually be functional and could just as easily associate multiple values with any input.
\{ \!\} expressions and ! clauses are especially expressive when used together in this way for direct recursion.

If we were to desugar the \{ \!\} syntax, ? clause, and ! clause in this rule, we would obtain two rules.
The rule below on the left emits a \rtag{do-subst} fact for the body of the lambda, if it qualifies for rewriting, and a \rtag{ruleXX-midpoint} fact saving pertinent details of the rule needed in its second half. Below on the right, the second half of the rule requires that the first half of the rule triggered and that the \rtag{subst} relation has responded with a rewritten lambda body for the \rtag{do-subst} fact \texttt{do'}.
\begin{multicols}{2}
\begin{Verbatim}[baselinestretch=0.8,commandchars=\\\{\}]
[(=/= x y)  ~(free E x)
 (= do (\rtag{do-subst} (\rtag{lam} x Ebody) y E)) 
 -->
 (= do' (\rtag{do-subst} Ebody y E)) 
 (\rtag{ruleXX-midpoint} do do' x)]
\columnbreak
[(\rtag{ruleXX-midpoint} do do' x)
 (\rtag{subst} do' Ebody')
 --> 
 (\rtag{subst} do (\rtag{lam} x Ebody'))]]
\end{Verbatim}
\end{multicols}
Finally, in the case of an application, substitution is performed down both subexpressions.
If one ! clause were nested under the other, they would need to be ordered. In this case, the compiler will detect that both !-clause facts can be emitted in parallel, so this rule will also split into two rules as in the rule above. The first rule will generate both !-clause facts and the second rule will await a response for both.
In this way, \slog{}'s semantics for ! clauses assists to naturally enable exposure of parallelism in semantics it models.

\begin{wrapfigure}{l}{6.5cm}
\begin{Verbatim}[baselinestretch=0.8,commandchars=\\\{\}]
(\rtag{subst} \huhclause{?(\rtag{do-subst} (\rtag{app} Ef Ea) x E)}
       (\rtag{app} \{\rtag{subst} \bangclause{!(\rtag{do-subst} Ef x E)}\}
	    \{\rtag{subst} \bangclause{!(\rtag{do-subst} Ea x E)}\}))
\end{Verbatim}
\end{wrapfigure}
With a substitution function defined, we can define evaluation using a pair of relations: \rtag{interp} and \rtag{do-interp}. A lambda \texttt{(\rtag{lam} x body)} is already fully reduced. An application reduces its left-hand subexpression to a lambda, substitutes the argument for the formal parameter, and reduces the body. 

\begin{wrapfigure}{r}{6cm}
\vspace{-0.15cm}
\begin{Verbatim}[baselinestretch=0.8,commandchars=\\\{\}]
\comm{; values}
(\rtag{interp} \huhclause{?(\rtag{do-interp} (\rtag{lam} x body))}
        (\rtag{lam} x body))

\comm{; application}
[(\rtag{interp} \bangclause{!(\rtag{do-interp} fun)} (\rtag{lam} x body))
 (\rtag{subst} \bangclause{!(\rtag{do-subst} body x arg)} body')
 -->
 (\rtag{interp} \huhclause{?(\rtag{do-interp} (\rtag{app} fun arg))}
         \{\rtag{interp} \bangclause{!(\rtag{do-interp} body')}\})]
\end{Verbatim}
\end{wrapfigure}
The compiler will detect in this case that the \rtag{do-subst} !-clause fact depends on variable \texttt{body}, and the \rtag{do-interp} ! clause in the head depends on \texttt{body'}, but that the first \rtag{do-interp} !-clause fact only depends on the variable fun from the original ?-clause fact kicking off the rule. These three sequential ! clauses split the rule into four parts during compilation, just as a continuation-passing-style (CPS) tranformation~\cite{appel2007compiling} would explicitly break a traditional functional implementation of this recursive, substitution-based interpreter into one function entry point and three continuation entry points. Unlike traditional CPS translation of functional programs however, ! clauses in \slog{} will naturally emit multiple facts for parallel processing in a nonblocking manner when variable dependence allows for parallelism. 

\subsection{Abstract Machines}
\label{sec:apps:am}
Next, instead of using terms alone to represent intermediate points in evaluation, we may wish to explicitly represent facets of evaluation such as the environment, the stack, and the heap. Instead of representing environments through substitution, we may want to represent them explicitly in a higher-order way. As shown in Figure~\ref{fig:defun-env}, with first-class facts and ad hoc polymorphic rules, we can use defunctionalization to implement first-class relations, providing a global \rtag{env-map} relation, we can read with an \texttt{\{\rtag{env-map} env x\}} expression (assuming ground variables \texttt{env} and \texttt{x}), along with a \texttt{(ext-env env x val)} facility for deriving an extended environment.
\begin{figure*}[h]
\begin{Verbatim}[baselinestretch=0.8,commandchars=\\\{\}]
\comm{; environments (defunctionalized)}
(\rtag{env-map} \huhclause{?(\rtag{ext-env} env x val)} x val)
[(=/= x y) --> (\rtag{env-map} \huhclause{?(\rtag{ext-env} env x _)} y \{\rtag{env-map} env y\})]
\end{Verbatim}
\caption{Defunctionalized environments; extension via \texttt{(ext-env env x v)}, lookup via \texttt{\{env-map env x\}}.}
\label{fig:defun-env}
\end{figure*}

On the left of Figure~\ref{fig:ce-machines} shows an abstract machine for CBN evaluation, and on the right, an abstract machine for CBV evaluation.
At the top, the rules for reference use \texttt{\{\rtag{env-map} env x\}} to access the value from the defunctionalized
\rtag{env-map} relation. In the CBN version, we cannot count on the stored closure to be a lambda closure, so we continue
interpretation, using another \{ \!\} expression to drop-in the transitive reduction of the stored argument closure.
Lambda closures are the base case which \rtag{interp} as themselves.
Finally, application closures trigger a closure to evaluate \texttt{Ef} via a ! clause, \texttt{\bangclause{!(clo Ef env)}}, and the lambda closure that finally results has its body evaluated under its environment, extended with parameter mapped to argument.
In the CBN interpreter, \texttt{(ext-env env' x (clo Ea env))} puts the argument expression \texttt{Ea} in the environment, closed with the current environment. In the CBV interpreter, \texttt{(ext-env env' x Eav)} puts the argument value \texttt{Eav} in the environment (after first evaluating it). In both these interpreters, the \rtag{app}-handling rules use ! clauses to implicitly create handling rules and a chain of continuation facts so \rtag{interp} maybe be utilized in a direct-recursive manner. The ! syntax introduces a CPS-like transformation that provides a stack in the interpretation of \slog{} rules for these CE interpreters to map their stack onto.
\begin{figure*}[h]
\begin{multicols}{2}
\vspace{-0.3cm}
\begin{Verbatim}[baselinestretch=.75,commandchars=\\\{\}]
\comm{; ref}
(\rtag{interp} \huhclause{?(\rtag{clo} (\rtag{ref} x) env)}
        \{\rtag{interp} \{\rtag{env-map} env x\}\})
\comm{; lam}
(\rtag{interp} \huhclause{?(\rtag{clo} (\rtag{lam} x Eb) env)}
        (\rtag{clo} (\rtag{lam} x Eb) env))
\comm{; app}
[(\rtag{interp} \bangclause{!(\rtag{clo} Ef env)}
         (\rtag{clo} (\rtag{lam} x Eb) env'))
 (= env'' (\rtag{ext-env} env' x (\rtag{clo} Ea env)))       
 (\rtag{interp} \bangclause{!(\rtag{clo} Eb env'')} v)
 -->
 (\rtag{interp} \huhclause{?(\rtag{clo} (\rtag{app} Ef Ea) env)} v)]  
\columnbreak
\comm{; ref }
(\rtag{interp} \huhclause{?(\rtag{clo} (\rtag{ref} x) env)}
        \{\rtag{env-map} env x\})
\comm{; lam}
(\rtag{interp} \huhclause{?(\rtag{clo} (\rtag{lam} x Eb) env)}
        (\rtag{clo} (\rtag{lam} x Eb) env))
\comm{; app}
[(\rtag{interp} \bangclause{!(\rtag{clo} Ef env)}
         (\rtag{clo} (\rtag{lam} x Eb) env'))
 (\rtag{interp} \bangclause{!(\rtag{clo} Ea env)} Eav)
 (\rtag{interp} \bangclause{!(\rtag{clo} Eb (\rtag{ext-env} env' x Eav))} v)
 -->
 (\rtag{interp} \huhclause{?(\rtag{clo} (\rtag{app} Ef Ea) env)} v)]
\end{Verbatim}
\end{multicols}
\caption{Two CE (closure-creating) interpreters in \slog{}; for CBN eval. (left) and CBV eval. (right).}
\label{fig:ce-machines}
\end{figure*}

\begin{figure*}[h]
\vspace{-0.75cm}
\hspace{-0.95cm}  
\begin{minipage}{0.98\linewidth}
\begin{multicols}{2}
\begin{Verbatim}[baselinestretch=.75,commandchars=\\\{\}]
    
\comm{; eval ref}
(\rtag{interp} \huhclause{?(\rtag{cek} (\rtag{clo} (\rtag{ref} x) env) k)}
        \{\rtag{interp} \bangclause{!(\rtag{cek} {\rtag{env-map} env x} k)}\})

      
\comm{; eval lam (apply)}
(\rtag{interp} \huhclause{?(\rtag{cek} (\rtag{clo} (\rtag{lam} x Eb) env)}
              \huhclause{[aclo k ...])}
        \{\rtag{interp} \bangclause{!(\rtag{cek} (\rtag{clo} Eb}
                       \bangclause{(\rtag{ext-env} env x aclo))}
                  \bangclause{k)}\})

                  

                  
\comm{; eval app}
(\rtag{interp} \huhclause{?(\rtag{cek} (\rtag{clo} (\rtag{app} Ef Ea) env) k)}
        \{\rtag{interp} \bangclause{!(\rtag{cek} (\rtag{clo} Ef env)}
                      \bangclause{[(\rtag{clo} Ea env) k ...])}\})
\comm{; return / halt}
(\rtag{interp} \huhclause{?(\rtag{cek} (\rtag{clo} (\rtag{lam} x Eb) env) [])}
        (\rtag{clo} (\rtag{lam} x Eb) env))

\columnbreak
\comm{; eval ref}
(\rtag{interp} \huhclause{?(\rtag{cek} (\rtag{clo} (\rtag{ref} x) env) k)}
        \{\rtag{interp} \bangclause{!(\rtag{cek} {\rtag{env-map} env x} k)}\})
\comm{; eval lam (ret to ar-k)}
(\rtag{interp} \huhclause{?(\rtag{cek} (\rtag{clo} (\rtag{lam} x Eb) env)}
              \huhclause{(\rtag{ar-k} aclo k))}
        \{\rtag{interp} \bangclause{!(\rtag{cek} aclo}
                      \bangclause{(\rtag{fn-k} (\rtag{clo} (\rtag{lam} x Eb) env)}
                            \bangclause{k))}\})
\comm{; eval lam (ret to fn-k)}
(\rtag{interp} \huhclause{?(\rtag{cek} (= aclo (\rtag{clo} (\rtag{lam} _ _) _))}
              \huhclause{(\rtag{fn-k} (\rtag{clo} (\rtag{lam} x Eb) env) k))}
        \{\rtag{interp} \bangclause{!(\rtag{cek} (\rtag{clo} Eb}
                           \bangclause{(\rtag{ext-env} env x aclo))}
                      \bangclause{k)}\})
\comm{; eval app}
(\rtag{interp} \huhclause{?(\rtag{cek} (\rtag{clo} (\rtag{app} Ef Ea) env) k)}
        \{\rtag{interp} \bangclause{!(\rtag{cek} (\rtag{clo} Ef env)}
                      \bangclause{(\rtag{ar-k} (\rtag{clo} Ea env) k))}\})
\comm{; return to (halt-k)}
(\rtag{interp} \huhclause{?(\rtag{cek} (\rtag{clo} (\rtag{lam} x Eb) env) (\rtag{halt-k}))}
        (\rtag{clo} (\rtag{lam} x Eb) env))
\end{Verbatim}
\end{multicols}
\end{minipage}
\caption{Two CEK (stack-passing) interpreters in \slog{}; for CBN eval. (left) and CBV eval. (right).}
\label{fig:cek-machines}
\end{figure*}

We can also implement the stack ourselves within our interpreter, thereby eliminating its need for our interpreter itself, by applying a stack-passing transformation.
On the left, Figure~\ref{fig:cek-machines} shows Krivine's machine~\cite{krivine:2007:cbn}, a tail-recursive abstract machine for CBN evaluation, and on the right, a tail-recursive abstract machine for CBV evaluation. Each of these machines incrementally constructs and passes a stack. In the CBN stack-passing interpreter, each application reached pushes a closure for the argument expression onto the stack. When a lambda is reached, this continuation is handled by popping its latest closure, the argument value. In the CBV stack-passing interpreter, each application reached pushes an \rtag{ar-k} continuation frame on the stack to save the argument value and environment. Whan a lambda is reached, this continuation is handled by swapping it for a \rtag{fn-k} continuation that saves the function value while the argument expression is evaluated (before application). Finally, when a lambda is reached, the \rtag{fn-k} continuation is handled by applying the saved closure. Now that the stack is entirely maintained by the interpreter itself, you may note that all recursive uses of \texttt{\{interp (cek ...)\}} are in tail position for the result column of relation \rtag{interp}.

\subsection{Abstracting Abstract Machines}
\label{sec:apps:aam}
\begin{wrapfigure}{r}{6.8cm}
\vspace{-0.75cm}
\begin{Verbatim}[baselinestretch=.75,commandchars=\\\{\}]
\comm{; eval ref -> ret}
[(\rtag{eval} (\rtag{ref} x) env sto k c)
 --> 
 (\rtag{ret} \{\rtag{sto-map} sto \{\rtag{env-map} env x\}\} sto k c)] 
\comm{; eval lam -> ret}
[(\rtag{eval} (\rtag{lam} x Eb) env sto k c)
 --> 
 (\rtag{ret} (\rtag{clo} (\rtag{lam} x Eb) env) sto k c)]
\comm{; eval app -> eval}
[(\rtag{eval} (\rtag{app} Ef Ea) env sto k c)
 --> 
 (\rtag{eval} Ef env sto (\rtag{ar-k} Ea env k) c)]
\comm{; ret to kaddr -> ret}
[(\rtag{ret} vf sto (\rtag{kaddr} c') c)
 -->
 (\rtag{ret} vf sto \{\rtag{sto-map} sto (\rtag{kaddr} c')\} c)]
\comm{; ret to ar-k -> eval}
[(\rtag{ret} vf sto (\rtag{ar-k} Ea env k) c)
 --> 
 (\rtag{eval} Ea env sto (\rtag{fn-k} vf k) c)]
\comm{; ret to fn-k -> apply}
[(\rtag{ret} va sto (\rtag{fn-k} vf k) c)
 --> 
 (\rtag{apply} vf va sto k c)]
\comm{; apply -> eval}
[(\rtag{apply} (\rtag{clo} (\rtag{lam} x Eb) env) va sto k c)
 --> 
 (\rtag{eval} Eb
       (\rtag{ext-env} env x (\rtag{addr} c))
       (\rtag{ext-sto} (\rtag{ext-sto} sto (\rtag{kaddr} c) k)
                (\rtag{addr} c) va)
       k
       \{+ 1 c\})]
\end{Verbatim}
\caption{A CESKT (control, environment, store, kontinuation, timestamp) interpreter in \slog{}.}
\label{fig:cesk-machine}
\vspace{-0.25cm}
\end{wrapfigure}

The \emph{abstracting abstract machines} (AAM) methodology \cite{might2010abstract,VanHorn:2010} proscribes a particular systematic application of abstract interpretation \cite{cousot77unifiedmodel,cousot1996abstract,cousot1979systematic} on abstract-machine operational semantics like those we've just built in \slog{}.
AAM proposes key preparatory refactorings of an abstract machine, to remove direct sources of unboundedness through recursion, before more straightforward structural abstraction can be applied.
In particular, there are two main sources of unboundedness in the CEK machines: environments and continuations. Environments contain closures which themselves contain environments; continuations are a stack of closures formed inductively in the CBV CEK machine and formed using \slog{}'s list syntax in the CBN CEK machine to more closely follow the usual presentation of Krivine's machine~\cite{krivine:2007:cbn}. 
AAM proposes threading each such fundamental source of unboundedness through a store, added in a normal store-passing transformation of the interpreter that might be used to add direct mutation or other effects to the language. Environments will map variables to addresses in the store, not to closures directly, and the stack will be store allocated at least once per function application so the stack may not grow indefinitely without the store likewise growing without bound. These two changes will permit us to place a bound on the addresses allocated, and thereby finitize the machine's state space as a whole.

Figure~\ref{fig:cesk-machine} shows the CBV CEK machine of Figure~\ref{fig:cek-machines} modified in a few key ways, yielding a CESKT machine with control expression, environment, store, continuation, and timestamp/contour components:
\begin{itemize}
\item
  \emph{abstract-machine states have been factored} into \rtag{eval}, \rtag{apply}, and \rtag{ret} configurations; an \rtag{eval} state has a control expression, environment (mapping variables to addresses), store (mapping addresses to closures and continuations), current continuation, and timestamp (tracking the size of the store, and thus the next address); an \texttt{apply} state has a closure being applied, argument value, store, continuation, and timestamp; and a \texttt{ret} state has a value being returned, a store, a continuation, and a timestamp;  
\item
  \emph{state transitions have been written as small-step rules} that always terminate; previously, our CEK machines were written to take a big-step from \rtag{cek}-state to the final, denoted value as logged in the \texttt{(\rtag{interp} e v)} relation, but in tail-recursive fashion, using ! clauses); Figure~\ref{fig:cesk-machine} has no explicit small-step relation, but simply says, for example, that the existence of a \rtag{ret} state permits us to deduce to existence of an \rtag{apply} state; if we were to want an explicit \rtag{step} relation, we could again give this rule a presentation with an implied body via a ? clause; for example:
\begin{Verbatim}[baselinestretch=.75,commandchars=\\\{\}]
(\rtag{step} \huhclause{?(\rtag{ret} va sto (\rtag{fn-k} vf k) c)}
      (\rtag{apply} vf va sto k c))
\end{Verbatim}
\item
  \emph{states have been subjected to a store-passing transformation} which has added a store \texttt{sto} and timestamp (stored-value count) \texttt{c} to each state; environments now bind variables to addresses and the current store binds those addresses to values; we perform a variable lookup with \texttt{\{\rtag{sto-map} sto \{\rtag{env-map} env x\}\}}; at an \rtag{apply} state, we use the store count \texttt{c} to generate a fresh address \texttt{(\rtag{addr} c)} for the parameter \texttt{x}; we also store-allocate the current continuation at a continuation address \texttt{(\rtag{kaddr} c)}, in preparation for modeling the stack finitely as well; when returning to a \texttt{(\rtag{kaddr} c)}, the continuation is simply fetched from the store as in the fourth rule down (ret to kaddr).
\end{itemize}

From here it suffices to pick a finite set from which to draw addresses. To instantiate a monovariant control-flow analysis from this CESKT interpreter, it would be enough to use the variable name itself as the address or to generate an address \texttt{(\rtag{addr} x)}. When the environment and store become finite, so does the number of possible states. Consider what happens, as the naturally relational \rtag{sto-map} relation encoding stores conflates multiple values at a single address for the same variable. Conflation in the store would lead naturally to nondeterminism in any \rtag{step} relation. When looking up a variable, two distinct \rtag{ret} states could result, leading to two distinct \rtag{apply} states after some further steps.

A (potentially) more precise, though (potentially) more costly analysis would be to specialize all control-flow points and store points by a finite history of recent or enclosing calls. Such a \emph{$k$-call-sensitive} analysis can be instantiated using a specific instrumentation and allocation policy, as can many others~\cite{gilray2016poly}. It requires an instrumentation to track a history of $k$ enclosing calls, and then an \emph{abstract allocation policy} that specializes variables by this call history at binding time. Such context-sensitive techniques are a gambit that the distinction drawn between variable \texttt{x} when bound at one call-site vs another will prove meaningful---in that it may correlate with its distinct values. Increasing the polyvariance allows for greater precision while also increasing the upper-bound on analysis cost. In a well known paradox of programming analyses, greater precision sometimes goes hand-in-hand with lower cost in practice because values that are simpler and fewer are simpler to represent~\cite{wright1998polymorphic}. At the same time, we use the polyvariant entry point of each function, its body and abstract contour---\texttt{(\rtag{kaddr} Eb c')}---to store allocate continuations as suggested by previous literature on selecting this address~\cite{gilray2016p4f} so as to adapt to the value polyvariance chosen.

The per-state store by itself is a source of exponential blowup for any polyvariant control-flow analysis~\cite{midtgaard2012control}. Instead, it is standard to use a global store and compute the least-upper-bound of all per-state stores. In \slog{} this is as simple as using a single global \texttt{(\rtag{store} addr val)} relation instead of a defunictionalized \texttt{(\rtag{sto-map} sto addr val)} relation that approximates all per-state stores in one. The left side of Figure~\ref{fig:kcfa-mcfa} shows a version of the CESKT machine with a global store and a tunable instrumentation that can be varied by changing the \rtag{tick} function rule. Currently, \rtag{tick} instantiates this to a $3$-$k$-CFA: at each function application, the current call site (now saved in the \rtag{ar-k} and \rtag{fn-k} continuation frames to provide to the apply state) is saved in front of the current call history and the fourth-oldest call is dropped.

This is the classic $k$-CFA, except perhaps that the original $k$-CFA, formulated for CPS as it was, also tracked returns positively instead of reverting the timestamp as functions return like we do here. The original $k$-CFA used true higher-order environments, unlike equivalent analyses written for object oriented languages which implicitly had flat environments (objects)~\cite{might2010resolving}. The corresponding CFA for functional languages is called $m$-CFA and is shown on the right side of Figure~\ref{fig:kcfa-mcfa}. $m$-CFA has only the latest call history as a flat context. Instead of having a per-variable address with a per-variable history tracked by a per-state environment, $m$-CFA stores a variable \texttt{x} at abstract contour \texttt{c} (i.e., abstract timestamp, instrumentation, 3-limited call-history) in the store at the address \texttt{(\rtag{addr} x c)}. This means at every update to the current flat context \texttt{c}, now taking the place of the environment, all free variables must be propagated into an address \texttt{(\rtag{addr} x c)}.

\begin{figure*}
%\vspace{-0.5cm}
\begin{multicols}{2}
\begin{Verbatim}[baselinestretch=.75,commandchars=\\\{\}]
\comm{;; Eval states}
[(\rtag{eval} (\rtag{ref} x) env k _)
 -->
 (\rtag{ret} \{\rtag{store} \{\rtag{env-map} env x\}\} k)]
[(\rtag{eval} (\rtag{lam} x body) env k _)
 -->
 (\rtag{ret} (\rtag{clo} (\rtag{lam} x body) env) k)]
[(\rtag{eval} (\rtag{app} ef ea) env k c)
 -->
 (\rtag{eval} ef env
         (\rtag{ar-k} ea env (\rtag{app} ef ea) c k)
         c)]
\comm{;; Ret states}
[(\rtag{ret} vf (\rtag{ar-k} ea env call c k))
 -->
 (\rtag{eval} ea env (\rtag{fn-k} vf call c k) c)]
[(\rtag{ret} va (\rtag{fn-k} vf call c k))
 -->
 (\rtag{apply} call vf va k c)]
[(\rtag{ret} v (\rtag{kaddr} e env))
 (\rtag{store} (\rtag{kaddr} e env) k)
  --> 
 (\rtag{ret} v k)]
\comm{;; Apply states}
[(\rtag{apply} call (\rtag{clo} (\rtag{lam} x Eb) env) va k c)
 -->
 (\rtag{eval} Eb env' (\rtag{kaddr} Eb env') c')
 (\rtag{store} (\rtag{kaddr} Eb env') k)
 (\rtag{store} (\rtag{addr} x c') va)
 (= env' (\rtag{ext-env} env x (\rtag{addr} x c')))
 (= c' \{\rtag{tick} \bangclause{!(\rtag{do-tick} call c)}\})]

\comm{;; tick (tuning for 3-k-CFA)}
(\rtag{tick} \huhclause{?(\rtag{do-tick} call [h0 h1 _])}
      [call h0 h1])

\columnbreak
\comm{;; Eval states}
[(\rtag{eval} (\rtag{ref} x) k c)
 -->
 (\rtag{ret} \{\rtag{store} (\rtag{addr} x c)\} k)]
[(\rtag{eval} (\rtag{lam} x body) k c)
 -->
 (\rtag{ret} (\rtag{clo} (\rtag{lam} x body) c) k)]
[(\rtag{eval} (\rtag{app} ef ea) k c)
 -->
 (\rtag{eval} ef (\rtag{ar-k} ea (\rtag{app} ef ea) c k) c)]


\comm{;; Ret states}
[(\rtag{ret} vf (\rtag{ar-k} ea call c k))
 -->
 (\rtag{eval} ea (\rtag{fn-k} vf call c k) c)]
[(\rtag{ret} va (\rtag{fn-k} vf call c k))
 -->
 (\rtag{apply} call vf va k c)]
[(\rtag{ret} v (\rtag{kaddr} e c))
 (\rtag{store} (\rtag{kaddr} e c) k)
 -->
 (\rtag{ret} v k)]
\comm{;; Apply states}
[(\rtag{apply} call (\rtag{clo} (\rtag{lam} x Eb) _) va k c)
 -->
 (\rtag{eval} Eb (\rtag{kaddr} Eb c') c')
 (\rtag{store} (\rtag{kaddr} Eb c') k)
 (\rtag{store} (\rtag{addr} x c') va)
 (= c' \{\rtag{tick} \bangclause{!(\rtag{do-tick} call c)}\})]
\comm{; Propagate free vars}
[(\rtag{free} y (lam x body))
 (\rtag{apply} call (\rtag{clo} (\rtag{lam} x body) clam) _ _ c)
 -->
 (\rtag{store} (\rtag{addr} y \{\rtag{tick} \bangclause{!(\rtag{do-tick} call c)}\})
        \{\rtag{store} (\rtag{addr} y clam)\})]
\end{Verbatim}
\end{multicols}
\caption{An AAM for global-store $k$-CFA (left) and $m$-CFA (right) in \slog{}. These are evaluated in Section~\ref{sec:eval}.}
\label{fig:kcfa-mcfa}
\end{figure*}

\subsection{Type Systems}
\label{sec:apps:ts}

Along with operational semantics and program analyses, \slog{}
naturally enables the realization of structural type systems based on
constructive logics~\cite{tapl,atapl}.  These systems are often
specified via an inductively-defined typing judgment, whose
derivations may be represented in \slog{} via (sub)facts and whose
typing rules may be realized as rules in \slog{} (providing their
evaluation may be operationalized via \slog{}'s idioms). For example,
the rules for the simply-typed $\lambda$-calculus (STLC) in
Figure~\ref{fig:stlc} define the judgment $\Gamma \vdash e
:\tau$---under typing environment $\Gamma$, $e$ has been proven to
have type $\tau$. Proofs of this judgment are represented via \slog{}
facts of the form \texttt{(\rtag{:} (\rtag{ck} $\Gamma$ e) T)}; each
rule in the type system is then mirrored by a corresponding rule in
\slog{}.

When implementing a type system in \slog{}, it is crucial to consider
some important differences between \slog{} and natural deduction
per-se. First, equivalence in \slog{} is intensional, via fact
interning (as in type theories such as Coq's Calculus of Inductive
Constructions). For example, while our type checker for STLC decides
$\Gamma \vdash e :\tau$ via structural recursion on $e$, there are
infinitely many $\Gamma' \supseteq \Gamma$ for which $\Gamma' \vdash e
:\tau$ also holds---materializing these (infinite) $\Gamma'$
would result in nontermination.

Here, we focus on the presentation of algorithmic (i.e.,
syntax-directed) type checking procedures. The decidability of these
systems follows immediately from their structurally-recursive nature,
a property inherited by their \slog{} counterparts. We expect
enumerating terms in other theories which enjoy strong normalization
will readily follow. We anticipate \slog{}'s declarative style may
also be a natural fit for type synthesis, by bounding (potentially
infinite) rewritings using a decreasing ``fuel'' parameter. However,
we leave this, along with explorations of other (bidirectional,
substructrural, etc...)  type systems in \slog{} to future work.

\paragraph*{Simply-typed $\lambda$-calculus}

\begin{figure}
\begin{displaymath}
\begin{tabular}{lrcllrcl}
\textit{STLC Terms}& $e$ & $::=$ & $\big(\lambda (x\!:\!\tau)\,e\big)$ & \textit{STLC Types}& $\tau \in T$ & $::=$& $\tau \rightarrow \tau$ \\
&     &  $\mid$ & $(e_0 ~ e_1)$ & && $\mid$& \textsf{nat} \\
&     &  $\mid$ & $x$ & & & $\mid$ & $...$ \\
\end{tabular}
\end{displaymath}

\begin{flushleft}
\fbox{$\Gamma \vdash e : \tau$}
\end{flushleft}
\begin{tabular}{c|c}
%\toprule \\
 {{\small\textsc{T-Var}}\quad\quad{\LARGE ${ \frac{x : T \, \in\, \Gamma}{\Gamma\, \vdash\, x : T}}$}} \quad&
\begin{minipage}{2.5in}
\begin{Verbatim}[baselinestretch=.8,commandchars=\\\{\},codes={\catcode`$=3\catcode`^=7}]
[-->;-------- T-Var \\
 (\rtag{:} \huhclause{?(\rtag{ck} $\Gamma$ (\rtag{ref} x))} \{\rtag{env-map} $\Gamma$ x\})]
\end{Verbatim}
\end{minipage}
\\
\\
{{\small\textsc{T-Abs}}\quad\quad{\LARGE ${ \frac{\Gamma, x\, : \,T_1 \, \vdash \, e \, : \, T_2}{(\lambda\,(x\,:\, T_1)\,e)\, : T_1 \rightarrow \,T_2}}$}} \quad& 
\begin{minipage}{2.5in}
\begin{Verbatim}[baselinestretch=.8,commandchars=\\\{\},codes={\catcode`$=3\catcode`^=7}]
[(\rtag{:} \bangclause{!(\rtag{ck} (\rtag{ext-env} $\Gamma$ x T1) e)} T2)
 -->;-------- T-Abs
 (\rtag{:} \huhclause{?(\rtag{ck} $\Gamma$ (\rtag{$\lambda$} x T1 e))} (\rtag{->} T1 T2))]
\end{Verbatim}
\end{minipage}
\\
\\
 
{{\small\textsc{T-App}}\quad\quad{\LARGE ${ \frac{{\Gamma \, \vdash \, e_0\,:\,T_0\, \rightarrow\, T_1}\quad e_1\,:\,T_0}{\Gamma \, \vdash \, (e_0~e_1)\,:\,T_1}}$ }}\quad& 
\begin{minipage}{2.5in}
\begin{Verbatim}[baselinestretch=.8,commandchars=\\\{\},codes={\catcode`$=3\catcode`^=7}]
[(\rtag{:} \bangclause{!(ck $\Gamma$ e0)} (\rtag{->} T0 T1))
 (\rtag{:} \bangclause{!(ck $\Gamma$ e1)} T0)
 -->;-------- T-App
 (\rtag{:} \huhclause{?(\rtag{ck} $\Gamma$ (\rtag{app} e0 e1))} T1)]
\end{Verbatim}
\end{minipage} \\
%\bottomrule
\end{tabular}

\caption{Syntax (top) and semantics (left) of STLC; equivalent \slog{} (right).}
\label{fig:stlc}
\end{figure}

We begin with the simply-typed
$\lambda$-calculus~\cite{barendregt2013lambda,tapl}.  The syntax of
STLC terms and types is shown at the top of Figure~\ref{fig:stlc}. Our
presentation roughly follows Chapter~9 of Benjamin Pierce's
\textit{Types and Programming Languages}~\cite{tapl}; we use
Scheme-style syntax to more closely mirror the \slog{}
presentation. STLC extends the untyped $\lambda$-calculus:
$\lambda$-abstractions are annotated with types, and variable typing
defers to a typing environment which assigns types to type variables
and is subsequently extended at callsites.  STLC defines a notion of
``simple'' types including (always) arrow types between simple types
and (sometimes) base types (e.g., \textsf{nat}), depending on the
presentation.
 
The key creative challenge in translating a type system into \slog{}
lies in the operationalization of its control flow. For example,
consider the \textsc{T-Var} rule on the left side of
Figure~\ref{fig:stlc}: the rule is parametric over the typing
environment $\Gamma$, however a terminating analysis necessarily
inspects only a finite number of typing environments. The solution is
interpret the rules in a demand-driven way, using subfacts to defer
type checking until necessary. Operationally, each type rule is
predicated upon a message \texttt{\huhclause{?(\rtag{ck} $\Gamma$
    e)}}, which triggers type synthesis for \texttt{e} under the
typing environment $\Gamma$.  Using this technique, we may invoke the
analysis by including a fact \texttt{(\rtag{ck} $\Gamma$ e)}. Using
stratified negation, we may treat \texttt{(\rtag{:} (\rtag{ck}
  $\Gamma$ e) $\tau$)} as a decision procedure, like so:

\begin{Verbatim}[baselinestretch=1,commandchars=\\\{\},codes={\catcode`$=3\catcode`^=7}]
[(\rtag{success} $\Gamma$ e $\tau$) <-- (\rtag{typecheck} $\Gamma$ e $\tau$) (\rtag{:} \bangclause{!(\rtag{ck} $\Gamma$ e)} $\tau$)]
[(\rtag{failure} $\Gamma$ e $\tau$) <-- ~(\rtag{success} $\Gamma$ e $\tau$)]
\end{Verbatim}

Here, the inclusion of \texttt{(\rtag{typecheck} $\Gamma$ e $\tau$)}
forces the materialization of \texttt{\bangclause{!(\rtag{ck} $\Gamma$
    e)}}, and subsequently the enumeration of the type of each of its
subexpressions. Once the resulting type is materialized, we force its
unification with $\tau$ (implicitly using \slog{}'s intensional notion
of equality) and, when successful, generate \texttt{(\rtag{success}
  $\Gamma$ e $\tau$)}. The resulting \slog{} implementation
necessarily terminates because the analysis enumerates at most a
finite number of \texttt{(\rtag{ck} $\Gamma$ e)} facts (because of the
structural recursion on $e$ done by \rtag{:}), which in-turn forces
materialization of a finite number of \texttt{(\rtag{ext-env}
  $\Gamma$ x T)} facts, each of which forces a bounded
materialization of \rtag{env-map}.

\paragraph*{Natural Deduction and Per Martin-L\"of's Type Theory}

The well-known Curry-Howard isomorphism relates terms in pure
functional languages to proofs in an appropriate constructive
logic~\cite{Curry:1934}. For STLC, the Curry-Howard isomorphism tells
us that we may read our type checker as a decision procedure for
intuitionistic propositional logic. As an alternative (but equivalent)
perspective, we now consider how \slog{} may represent proofs in Per
Martin-L\"of's intuitionistic type theory (ITT)~\cite{Martin-Lof1996}.

By design, ITT cleanly separates propositions from their associated
derivations (proof objects). In \slog{}, we may represent derivations
as structured facts, obtaining the nested structure of derivations via
\slog{}'s subfacts. Adopting this perspective, checking a
natural-deduction proof in ITT involves propagating assumptions to
their usages (akin to the propagation achieved using maps in STLC).

\begin{figure}[h!]

\begin{tabular}{c|c}
{{\textsc{$\land$I}}\quad{\LARGE  ${ \frac{A \textit{ true} \quad \quad B \textit{ true}}{A \land B \textit{ true}}}$}} \quad& 
\begin{minipage}{2.5in}
{
\begin{Verbatim}[baselinestretch=.8,commandchars=\\\{\},codes={\catcode`$=3\catcode`^=7}]
[(\rtag{true} \bangclause{!(\rtag{ck} A-pf A)})
 (\rtag{true} \bangclause{!(\rtag{ck} B-pf B)})
 -->
 (\rtag{true} \huhclause{?(\rtag{ck} (\rtag{$\land$I} A-pf B-pf (\rtag{$\land$I} A B)) (\rtag{$\land$I} A B))})]
\end{Verbatim}
}
\end{minipage}
\end{tabular}
\caption{The introduction rule for $\land$ in ITT.}
\label{fig:itt}
\end{figure}

Figure~\ref{fig:itt} shows the introduction form for $\land$
(left) and its corresponding \slog{} transliteration (right). Checking
a derivation of $\land$I forces the checking of each sub-derivation,
and (upon success) populates the \rtag{true} relation with the
appropriate derivation.

Implication in ITT is managed by introducing, and then discharging,
assumptions: $A \supset B$ holds whenever $B$ may be proven by
assuming $A$. Figure~\ref{fig:impl} details the implication rule in
ITT (left) and \slog{} (right): the introduced hypothesis (named $u$
in $\supset{}\!I$) is subsequently discharged to produce an
assumption-free proof of $A \supset B~\textit{true}$. In \slog{}, the
rule for $\supset{}\!I$ introduces an assumption by forcing the
materialization of \texttt{(\rtag{assuming} A pf-B)}---other rules
then ``push down'' the assumption $A~\textit{true}$ to their eventual
uses (top right of Figure~\ref{fig:impl}), performing transitive
closure of assumptions to their usages in an on-demand fashion.

\begin{figure}[h!]
\begin{tabular}{c|c}
%% -- u
%% A  
%% ..
%% B 
%% -- 
%% A -> B

${\supset{}\!I}\quad \frac { \frac{}{\begin{array}{c}A~\textit{true}\\\vdots\\ B~\textit{true}\end{array}}~u } {\begin{array}{c}A \supset B~\textit{true}\end{array}}$

&
\begin{minipage}{2.5in}
{
\vspace{-.5in}
\begin{Verbatim}[baselinestretch=.8,commandchars=\\\{\},codes={\catcode`$=3\catcode`^=7}]
;; Propagate assumptions
(\rtag{true} \huhclause{?(\rtag{ck} (\rtag{assuming} P (\rtag{assumption} P)) P)} P)
...

[(\rtag{true} \bangclause{!(\rtag{ck} (\rtag{assuming} A pf-B) B)})
 -->
 (\rtag{true} \huhclause{?(\rtag{ck} (\rtag{$\supset$I} (\rtag{$\supset$} A B) pf-B) (\rtag{$\supset$} A B))})]
\end{Verbatim}
}
\end{minipage}
\end{tabular}
\caption{Implication in ITT and \slog{}}
\label{fig:impl}
\end{figure}

\paragraph*{First-Order Dependent Types: \lf{}}

\begin{figure}
\begin{displaymath}
\begin{tabular}{lrcllrcl}
\textit{\lf{} Contexts}& $\Gamma$ & $::=$ & $\varnothing$ &\textit{\lf{} Types} & $\tau \in T$ & $::=$& $X$ \\
&     &  $\mid$ & $\Gamma, x \! : \! T$ & && $\mid$& $\Pi x \! : \! T.~ T$ \\
&     &  $\mid$ & $\Gamma, x \! ::\!  K$ & &&$\mid$& $(T~ e)$ \\
\textit{\lf{} Kinds} & $K$ & $::=$ & $\Pi x \! : \! T. K \mid \ast$ \\ 
\end{tabular}
\end{displaymath}
\begin{tabular}{c|c}
%\toprule \\
{{\small\textsc{TA-Abs}}\quad\quad{\LARGE ${ \frac{\Gamma\, \vdash\, S\, ::\, \ast \quad \Gamma, \,x\, :\, S\, \vdash\, t\, : \,T}{\Gamma\, \vdash \, (\lambda (x\,:\,S) t) \,:\, \Pi x\, :\, S. T}}$}} \quad& 
\begin{minipage}{2.5in}
{\footnotesize
\begin{Verbatim}[baselinestretch=.8,commandchars=\\\{\},codes={\catcode`$=3\catcode`^=7}]
[(\rtag{::} \bangclause{!(\rtag{ck-k} $\Gamma$ S)} (\rtag{$\ast$}))
 (\rtag{:} \bangclause{!(\rtag{ck-t} (ext-env $\Gamma$ x S) t)} T)
 -->;------
 (\rtag{:} \bangclause{?(\rtag{ck-t} $\Gamma$ (\rtag{$\lambda$} x S t))} (\rtag{$\Pi$} x S T))]
\end{Verbatim}
}\end{minipage}
\\
\\
  {{\small\textsc{TA-App}}\quad\quad{\LARGE ${ \frac{\Gamma \, \vdash \, t_1 \,:\, \Pi x : S_1 . \, T \quad \Gamma \, \vdash \, t_2\, : \,S_2 \quad \Gamma\, \vdash\, S_1\, \equiv\, S_2}{\Gamma\, \vdash\, (t_1~t_2)\, : \,[ x \,\mapsto\, t_2 ] \, T}}$}}\quad&
\begin{minipage}{2.5in}
{\footnotesize
\begin{Verbatim}[baselinestretch=.8,commandchars=\\\{\},codes={\catcode`$=3\catcode`^=7}]
[(\rtag{:} \bangclause{!(\rtag{ck-t} $\Gamma$ t1)} (\rtag{$\Pi$} x S1 T))
 (\rtag{:} \bangclause{!(\rtag{ck-t} $\Gamma$ t2)} S2)
 (\rtag{true} \bangclause{!(\rtag{===} $\Gamma$ S1 S2)})
 -->;------
 (\rtag{:} \huhclause{?(\rtag{ck-t} $\Gamma$ (\rtag{app} t1 t2))}
   \{\rtag{subst} \bangclause{!(\rtag{do-subst} T x t2)}\})]
\end{Verbatim}
}\end{minipage} \\
\\
  {{\small\textsc{KA-App}}\quad\quad{\LARGE ${ \frac{\Gamma \, \vdash \, S \,::\, \Pi x : T_1 . \, K \quad \Gamma \, \vdash \, t\, : \,T_2 \quad \Gamma\, \vdash\, T_1\, \equiv\, T_2}{\Gamma\, \vdash\, (S~t)\, : \,[ x \,\mapsto\, t ] \, K}}$}}\quad&
\begin{minipage}{2.5in}
{\footnotesize
\begin{Verbatim}[baselinestretch=.8,commandchars=\\\{\},codes={\catcode`$=3\catcode`^=7}]
[(\rtag{::} \bangclause{!(\rtag{ch-k} $\Gamma$ S)} (\rtag{$\Pi$} x T1 K))
 (\rtag{:} \bangclause{!(\rtag{ck-t} $\Gamma$ t)} T2)
 (\rtag{true} \bangclause{!(\rtag{===} $\Gamma$ T1 T2)})
 -->;------
 (\rtag{:} \huhclause{?(\rtag{ch-t} $\Gamma$ (\rtag{type-app} S t))}
   \{subst !(do-subst K x t)\})]
\end{Verbatim}
}\end{minipage} \\
\\
%\bottomrule
\begin{minipage}{2.75in}
{\small
\begin{Verbatim}[baselinestretch=.8,commandchars=\\\{\},codes={\catcode`$=3\catcode`^=7}]
[(\rtag{->wh} \bangclause{!(\rtag{do->wh} t1)} t1')
 -->;------
 (\rtag{->wh} \huhclause{?(\rtag{do->wh} (\rtag{app} t1 t2))} (\rtag{app} t1' t2))]
\end{Verbatim}
}
\end{minipage}
&
\quad
\quad
\begin{minipage}{2.75in}
{\small
\begin{Verbatim}[baselinestretch=.8,commandchars=\\\{\},codes={\catcode`$=3\catcode`^=7}]
(\rtag{->wh} \huhclause{?(\rtag{do->wh} (\rtag{app} ($\lambda$ x T1 t1) t2))}
  \{subst \bangclause{!(do-subst t1 x t2)}\})
\end{Verbatim}
}\end{minipage} 
\end{tabular}

\caption{\lf{}: Syntax (top), selected rules (left) and \slog{} (right).}
\label{fig:lf}
\end{figure}

The Edinburgh Logical Framework (\lf{}) is a dependently-typed
$\lambda$-calculus~\cite{Harper:93}. It is a first-order dependent
type system, in the sense that it stratifies its objects into kinds, types
(families), and terms (values)---kinds may quantify over types, but
% Todo: Why put this idea here right up front?
not over other kinds. The syntax of \lf{} is detailed at the bottom of
Figure~\ref{fig:lf}---it extends STLC with kinds, which are either
$\ast$ or (type families) $\Pi x : T. K$, where $T$ is a simple type
(of kind $\ast$). \lf{} generalizes the arrow type to the dependent
product: $\Pi x: T. T$. System \lf{} enjoys several decidability
properties which make it particularly amenable to implementation in
\slog{}. The first is strong normalization, which implies that
reduction sequences for well-typed terms in our implementation will be
finite. The second is \lf{}'s focus on canonical forms and hereditary
substitution~\cite{harper:2007}. In \lf{}, terms are canonicalized to
weak-head normal form (WHNF); this choice enables inductive reasoning
on these canonical forms, and this methodology forms the basis for
% Todo: Say a bit more about Twelf? It's meant for theorem proving or programming?
Twelf~\cite{pfenning1998twelf}.

The \textit{judgments-as-types} principle interprets type checking for
\lf{} as proving theorems in intuitionstic predicate logic; Using this
principle, we may define traditional constructive connectives (such as
$\land$, $\lor$, and $\exists$) via type families and their associated
rules. For example, including in $\Gamma$ a binding $\land \mapsto
\Pi\, P : \textit{prop}.~ \Pi\, Q : \textit{prop}.~ \ast$ allows using
the constructor $\land$, though $\land$ must be instantiated with a
suitable $P$ and $Q$, which must necessarily be of some sort (e.g.,
\textit{prop}) also bound in $\Gamma$.

We have performed a transliteration of \lf{} as formalized in Chapter
2 of \textit{Advanced Topics in Types and Programming Languages
  (ATAPL)}~\cite{atapl}. Our transliteration (from pages 57--58)
consists of roughly 150 lines of \slog{}
code. Figure~\ref{fig:lf} details several of the key
rules. \textsc{TA-Abs} introduces a $\Pi$ type, generalizing the
\textsc{T-Abs} rule in Figure~\ref{fig:stlc}. The \textsc{TA-App}
applies a term $t_1$, of a dependent product type $\Pi x :S_1 ~.~T$,
whenever the input $t_2$ shares an equivalent type, $S_2$. The notion
of equality here is worth mentioning: $\equiv$ demands reduction of
its arguments to WHNF---\lf{} is constructed to identify terms under
WHNF, thus ensuring $\equiv$ will terminate as long as the term is
typeable. Reduction to WHNF is readily implemented in \slog{}; two
exemplary rules detailed at the bottom of Figure~\ref{fig:lf}
% Todo: I changed this from fig:lf-example; correct?
outline the key invariant in WHNF: reduce down the leftmost spine,
eliminating $\beta$-redexes via application. Equality checks are
demanded by the \textsc{TA-App} and \textsc{KA-App} rules, and force
normalization of their arguments to WHNF before comparison of
canonical forms, generating a witness in \rtag{true} before triggering
the head of the rule.

%% \begin{tabular}{c|c|c}
%% \begin{align}
%% \forall \, Q,&\, \forall\, P.& \\
%%   &\,Q \land P \rightarrow P \land Q
%% \end{align}
%% &
%% \begin{minipage}{2.75in}
%% {\small
%% \begin{Verbatim}[baselinestretch=.6,commandchars=\\\{\},codes={\catcode`$=3\catcode`^=7}]
%% (tcheck
%%  (extend
%%   (extend (bot) (var (prop)) (star))
%%   (var (land)) (PiK (x) (var (prop))
%%                     (PiK (y) (var (prop)) (star))))
%%  (lam (x) (var (prop))
%%       (lam (y) (var (prop))
%%            (type-app (var (land)) (ref (x))))))
%% \end{Verbatim}
%% }
%% \end{minipage}
%% &
%% \quad
%% \quad
%% \begin{minipage}{2.75in}
%% {\small
%% \begin{Verbatim}[baselinestretch=.6,commandchars=\\\{\},codes={\catcode`$=3\catcode`^=7}]
%% (tcheck
%%  (extend
%%   (extend (bot) (var (prop)) (star))
%%   (var (land)) (PiK (x) (var (prop))
%%                     (PiK (y) (var (prop)) (star))))
%%  (lam (x) (var (prop))
%%       (lam (y) (var (prop))
%%            (type-app (var (land)) (ref (x))))))
%% \end{Verbatim}
%% }\end{minipage} 
%% \end{tabular}

\section{Implementation}

We have implemented \slog{} in a combination of Racket (the compiler,
roughly 10,600 lines), \CC{} (the runtime system and parallel RA
backend; roughly 8,500 lines), Python (a REPL and daemon; roughly
2,500 lines) and Slog (60 lines for list-splicing support). In this
section, we describe relevant particulars.

\subsection{Compiler}

Our Racket-based compiler translates Slog source code to \CC{} code
that links against our parallel RA backend. Our compiler is structured
in the nanopass style, composed of multiple passes which consume and
produce various forms of increasingly-low-level intermediate
representations (IRs)~\cite{nanopass}. After parsing,
\emph{organize-pass} performs various simplifications, such as
canonicalizing the direction of \texttt{-\ \!\!\!\!->} and splitting disjunctive
body clauses and conjunctive head clauses into multiple rules.
This pass also eliminates various syntactic niceties of the language, including \text{!} and \text{?}
clauses, nested rules, list syntax, and splicing variables. When a
program includes splicing variables, the compiler concatenates a small
splicing support library. Finally, this pass performs static
unification, syntactically identifying clauses and variables that are
statically constrained to be equal.

\slog{}'s distribution paradigm is built upon
binary joins. Thus, after organization, \emph{partitioning-pass}
breaks down bodies with multiple clauses into sequences of binary
joins. Partitioning represents an algorithmic challenge, as there may
be many ways to partition a set of clauses into sequences of binary
joins. For example, a rule such as
\lstinline[mathescape]{[H <-- B$_1$ B$_2$ $...$ B$_n$]}
may be converted into an $(n-1)$-length sequence of
binary joins (first joining $B_1$ and $B_2$ to form an intermediate
relation which is subsequently joined with $B_3$) or a tree of binary
joins (joining each $B_i$ and $B_{i+1}$ into intermediate relations
which are then joined). Unfortunately, optimal partitioning is
undecidable in general, and our compiler relies upon a set of
heuristics and practical optimizations we've found to work well in practice, along
with enabling the user to manually suggest a partitioning by using
a \lstinline{--} syntax operator. Our early implementations preferred
to form trees of joins---based on the intuition that this would extract
more parallel work---however, we found this often resulted in materializing
large numbers of unnecessary facts (essentially forming large Cartesian
products to later be filtered in subsequent joins). We have come to
see effective partitioning as a fundamental part of writing
high-performance code in \slog{}, similar to Souffl\'e's Sideways
Information Passing Strategy~\cite{soufflesips}.

After partitioning, \emph{split-selections-pass} performs index
selection by inspecting each rule in the program and calculating a set
of indices. There are two important differences between \slog{} and a
typical shared-memory Datalog implementation. First, because of our
distribution methodology, it is impossible to organize indices in a
trie-shaped representation that allows overlap-based index
compression~\cite{Subotic:2018:AIS:3282495.3302538}. Second, fact
interning is slightly tricky in the presence of multiple indices: we
must be careful to avoid a fact being assigned two distinct intern
keys during the same iteration in separate indices. Our solution is to
designate a special \emph{canonical index} which is used for intern
key origination, along with a set of special administrative rules which
replicate, for each relation, the intern key to every non-canonical
index. 

The last two passes are strongly-connected component (SCC)
construction and incrementalization. Datalog programs are typically
stratified into a plan of SCCs for evaluation. This aids efficiency (a
single-node implementation may ignore considering rules unnecessary to
the current SCC, our distributed implementation evaluates SCCs using
task-level parallelism) and is also semantically relevant in the
presence of stratified negation (to ensure all negated relations are
computed strictly-before negation runs). Finally, incrementalization
(i.e., semi-na\"ive evaluation) transforms the program to use a
worklist-based evaluation strategy, every relation appearing in a rule
body is split into two versions---\textbf{delta} and
\textbf{total}. Rules are rewritten to add facts to \textbf{delta},
while bodies are triggered by new entries in \textbf{delta}; our
backend merges \textbf{delta} into \textbf{total} at the end of each
iteration.

\subsection{Backend}

\begin{figure*}[t]
\includegraphics[width=0.99\textwidth]{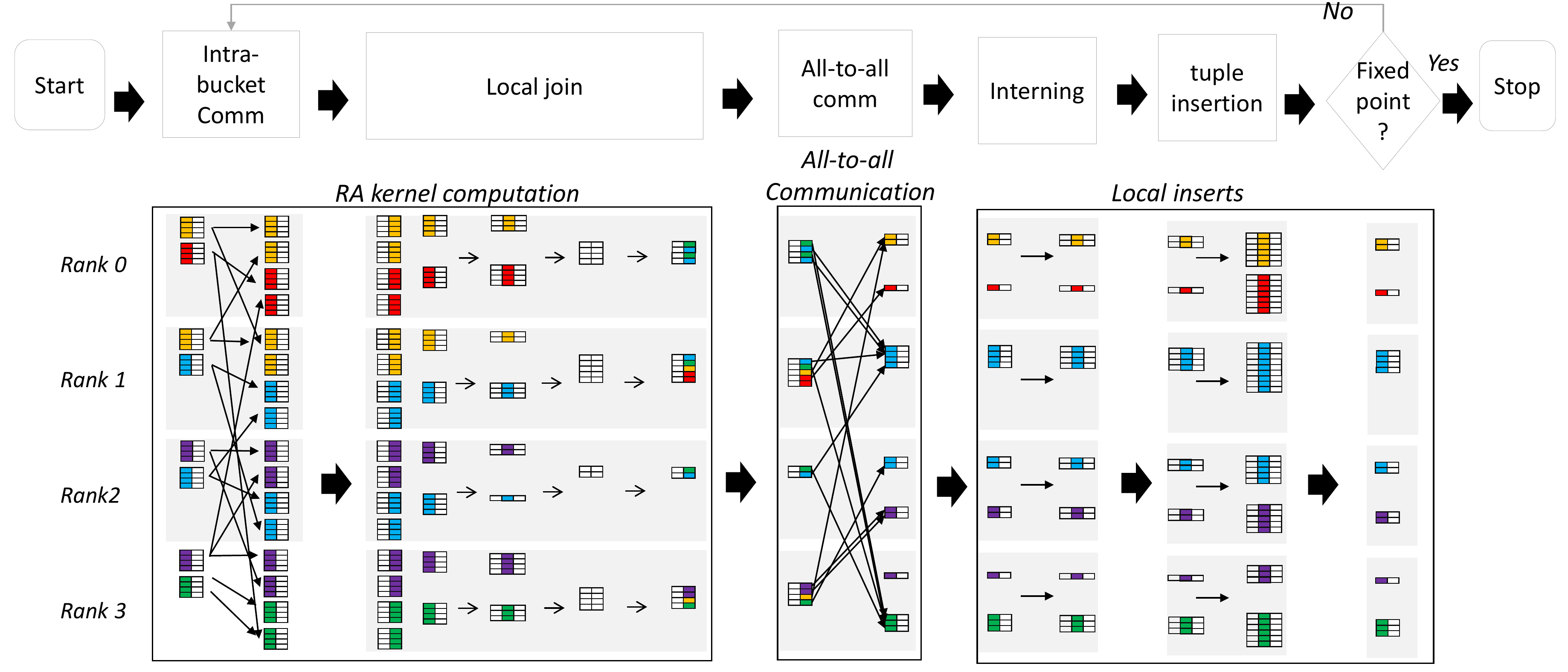}
\caption{An illustration of the main phases of our parallel RA backend.}
\label{fig:sid}
\end{figure*}

Our parallel relational-algebra backend supports fixed-point
iterations and is designed for large-scale multi-node HPC
clusters. Based on the bulk-synchronous-processing protocol and built
using the \texttt{MPI-everywhere} model~\cite{forum1994mpi, zambre2021logically},
the parallel RA framework addresses the problem of partitioning and balancing workloads across
processes by using a two-layered distribution
hash-table~\cite{kumar:hipc:2019}. In order to materialize newly generated facts within each iteration, and thus facilitate \emph{iterated} RA (in a fixed-point loop), an all-to-all data exchange phase is used at every iteration. Figure~\ref{fig:sid} shows a schematic diagram of all the phases (including the interning phase) in the context of an incrementalized TC computation. There are three primary phases in the backend: (1) RA kernel computation, (2) all-to-all communication and (3) local insertion.

Figure~\ref{fig:sid} shows a schematic diagram of all the phases (including the interning phase) in the context of an incrementalized TC computation. There are three primary phases of our system: local RA computation, all-to-all data exchange, and materialization in appropriate indices. Workload (relations) are partitioned across processes using the the double hashing approach, which also adds an extra intra-bucket communication phase, to co-locate matching tuples. During the local computation phase, RA kernels (comprising of join, projection, union and others) are executed in parallel across all processes. Output tuples generated from local computation may each belong to an arbitrary bucket in the output relation, so an MPI all-to-all communication phase shuffles the output of all joins to their managing processes (preparing them for any subsequent rules or iterations). Once new tuples are received, we perform the interning phase, that first checks if the received fact was already populated before, if not then a new intern id is created and associated with the fact. The fact is then inserted into newt, and following the semi-naive evaluation approach, the newly generated facts forms the input for the following iteration of the fixed point. This process continues till the fixed-point is reached and no new facts can be discovered.

\paragraph*{RA kernel computation}
The two-layered distributed approach, with local hash-based joins and hash-based distribution of relations, is a foundational method to distribute RA-kernel (primarily join) operations over many nodes in a networked cluster computer. This algorithm involves partitioning relations by their join-column values so that they can be efficiently distributed to participating processes~\cite{Valduriez:1988:PET:54616.54618}. The main insight behind this approach is that for each tuple in the outer relation, all relevant tuples in the inner relation must be hashed to the same MPI process or node, permitting joins to be performed locally on each process.

A major challenge with parallel workload partitioning is to ensure that every process gets to work on similar sized workloads that gives a load balanced system. A major challenge to enforce this load balance is to deal with inherently imbalanced data coming from key-skewed relations. To ensure uniform load across processes, we have built on previous approaches~\cite{kumar:hipc:2019, Kumar:2020} by developing strategies that mitigate load-imbalance in a dynamic manner. 
%First, we describe the architecture of our join (synthesized with projection and renaming for TC computation) in detail to ground this discussion. 
The approach~\cite{kumar:hipc:2019} uses a two-layered distributed hash-table to partition tuples over a fixed set of \emph{buckets}, and, within each bucket, to a dynamic set of \emph{subbuckets} which may vary across buckets. Each tuple is assigned to a bucket based on a hash of its key-column values, but within each bucket tuples are hashed on non-join-column values, assigning them to a local subbucket, then mapped to an MPI process. Within subbuckets, tuples are stored in B-trees, organized by key-column values. Our scheme permits buckets that have more tuples to be split across multiple processes, but requires some additional communication among subbuckets for any particular bucket. We have developed and evaluated a dynamic refinement strategy, to decide how many subbuckets to allocate per-bucket. To distribute subbuckets to managing processes, we use a round-robin mapping scheme which we found to be significantly more effective than hashing.

A join operation can only be performed for two \emph{co-located relations}: two relations each keyed on their respective join columns that share a bucket decomposition (but not necessarily a subbucket decomposition for each bucket). This ensures that the join operation may be performed separately on each bucket as all matching tuples will share a logical bucket; it does not, however, ensure that all two matching tuples will share the same subbucket as tuples are assigned to subbuckets (within a bucket) based on the values of non-join columns, separately for each relation.
The first step in a join operation is therefore an \emph{intra-bucket communication} (see Figure~\ref{fig:sid}) phase within each bucket so that every subbucket receives all tuples for the outer relation across all subbuckets (while the inner relation only needs tuples belonging to the local subbucket). Following this, a \emph{local join} operation (with any necessary projection and renaming) is performed in every subbucket.

\paragraph*{All-to-all communication}
To enable iterated parallel RA (in a fixed-point loop), processes must engage in a non-uniform all-to-all inter-process shuffle of generated tuples to their position in an output index.
This data exchange is performed to \emph{materialize} the output tuples generated from the local compute phase (where RA kernels are executed) to their appropriate processes (based on their bucket-subbucket assignment).
Materializing a tuple in an output relation (resulting from an RA operation) involves hashing on its join and non-join columns to find its bucket and sub-bucket (respectively), and then transmitting it to the process that maintains that bucket/sub-bucket.
As tuples generated from the local compute phase may each belong to an arbitrary bucket/sub-bucket in the output relation, an all-to-all communication phase is required to shuffle the output tuples to their managing processes.
Given variability in the number of tuples generated across processes, and in their destination processes (due to inherent imbalance), the communication phase in our framework is \emph{non-uniform} in nature.
The output tuples may each belong to an arbitrary bucket in the output relation, an MPI \emph{all-to-all} communication phase shuffles the output of all joins to their managing processes (preparing them for any subsequent iteration).

The overall scalability of the RA backend relies on the scalability of the all-to-all inter-process data exchange phase.
However, all-to-all is notoriously difficult to scale~\cite{4536141, scott1991efficient, thakur2005optimization}---largely because of the \emph{quadratic} nature of its workload. 
We address this scaling issue by adopting recent advancements~\cite{fan2022optimizing} that optimizes non-uniform all-to-all communication by extending the $\log$-time Bruck algorithm~\cite{bruck1997efficient, thakur2005optimization, traff2014implementing} for non-uniform all-to-all workloads. Traditional algorithms to implement non-uniform all-to-all communication takes linear iterations as every process must send and receive from every other process. Bruck algorithm however, sends more amount of data in logarithmic steps, and therefore significantly improves overall performance of all-to-all data exchange. Using the Bruck implementation of non-uniform all-to-all algorithm was instrumental in the overall scalibility of the backend.

\paragraph*{Local inserts}
After all-to-all data exchange, every process receives a new set of facts that must be materialized to be used as input in the subsequent iteration of the fixed-point loop. Local insert is a two-step process involving interning and inserting newly generated facts in the appropriate version of a relation (delta and total). Interning assigns a unique 64-bit key to every fact, and in order to scale this process, it must be is performed in an embarrassingly parallel manner without the need for any synchronization among processes. This is done by reserving the first 16 bits of the key for unique buckets ids, and the remaining 48-bits for facts. Since, a canonical-index (master index) bucket is never split across a process, reserving 16 bits for bucket ids ensures that globally unique intern keys can be created concurrently across processes. The fact-id component of the intern key is created by a bump pointer, which ensures that locally all facts receive a unique key. After interning, facts are added to their appropriate versions of the relations (delta or total), and a check is performed to see if a fixed-point has been reached. This check is performed by a global operation that checks the size of all relations across all processes, and if all sizes remains unchanged across two iterations, then this indicates that fixed-point has been attained and the program terminates, otherwise a new iteration is initiated.

\section{Evaluation}
\label{sec:eval}

We aimed to measure and evaluate \slog{}'s
improved indexing and data parallelism, using
three sets of performance benchmarks (PBs):

\begin{description}
\item[\textbf{PB1}] (Section~\ref{sec:eval:pb1}) How does \slog{} compare against other systems
  designed for performance and parallelism on traditional Datalog
  workloads (without ADTs): Souffl\'e and RadLog?
\item[\textbf{PB2}] (Section~\ref{sec:eval:pb2}) How do \slog{} subfacts perform against Souffl\'e ADTs
  in the context of the $m$-CFA and $k$-CFA benchmarks developed in Section~4.
\item[\textbf{PB3}] (Section~\ref{sec:eval:pb3}) How well can \slog{} scale to many threads on a supercomputer?
\end{description}

We evaluated \textbf{PB1} and \textbf{PB2} by running a set of
experiments on large cloud machines from Amazon AWS and Microsoft
Azure. For \textbf{PB1}, we ran a set of strong scaling experiments of
transitive closure on large graphs, picking transitive closure as an
exemplary problem to measure end-to-end throughput of deductive
inference at scale. For \textbf{PB2}, we measure the performance of
the implementation of our $k$ and $m$-CFA analyses from
Section~\ref{sec:apps} compared to an equivalent implementation in
Souffl\'e using abstract datatypes (ADTs). We answer \textbf{PB3} by
running experiments on the Theta supercomputer at Argonne National
Supercomputing Lab, scaling a control-flow analysis for the
$\lambda$-calculus to $1000$ threads on Argonne's Theta.

\subsection{Transitive Closure}
\label{sec:eval:tc}
\label{sec:eval:pb1}

We sought to compare \slog{}'s full-system throughput on vanilla
Datalog against two comparable production systems: Souffl\'e and
Radlog. Souffl\'e is engineered to achieve the best-known performance
on unified-memory architectures, and supports parallelism via
OpenMP. Radlog is a Hadoop-based successor to the BigDatalog deductive
inference system, which uses Apache Spark to perform distributed joins
at scale~\cite{rasql}. We originally sought to compare \slog{}
directly against BigDatalog, but found it does not support recent
versions of either Spark or Java (being built to target Java
1.5). Under direction of BigDatalog's authors, we instead used Radlog,
which is currently under active development and runs on current
versions of Apache Spark.

We performed comparisons on an \textsf{Standard\_M128s} instance
rented from Microsoft Azure~\cite{mseries}. The node used in our
experiments has 64 physical cores (128 threads) running Intel Xeon
processors with a base clock speed of 2.7GHz and 2,048GB of RAM. To
directly compare \slog{}, Souffl\'e, and Radlog, we ran each on the
same machine using 15, 30, 60, and 120 threads. We ran \slog{} using
OpenMPI version 4.1.1 and controlled core counts via
\texttt{mpirun}. We compiled Souffl\'e from its Git repository, using
Souffl\'e's compiled mode to compile each benchmark separately to use
the requisite number of threads before execution. Radlog runs natively
on Apache Spark, which subsequently runs on Hadoop. To achieve a fair
comparison against Souffl\'e and \slog{}, we ran Radlog using Apache
Spark configured in local mode; Spark's local mode circumvents the
network stack and runs the application directly in the JVM. We used
three large graphs shown the first column of Table~\ref{tab:single-results}:
\textsc{fb-media} is media-related pages on Facebook,
\textsc{ring10000} is ring graph of 10,000 nodes, and
\textsc{suitesparse} is from the UF Sparse Matrix
Collection~\cite{Davis:2011:UFS:2049662.2049663}. We configured Radlog
according the directions on its website, experimenting with a variety
of partitions (used for shuffling data between phases) to achieve the
best performance we could. Ultimately, we used three times as many
partitions as available threads, except for \textsc{ring10000}, for
which we found higher partition counts caused significantly lower
performance.

\begin{table}
\centering
\caption{Single-node TC Experiments}
\begin{tabular}{m{1.7cm}m{1cm}m{2cm}|cm{.5cm}m{.5cm}m{.5cm}m{.5cm}}
\toprule
\multicolumn{3}{c}{Graph Properties}&& \multicolumn{4}{c}{Time (s) at Process Count} \\
Name & Edges & {\quad$\mid$TC$\mid$} & System & 15 & 30 & 60 & 120 \\
\midrule
\multirow{3}{*}{\textsc{fb-media}} & \multirow{3}{*}{206k} & \multirow{3}{*}{96,652,228} & \slog & 62 & 40 & 21 & \textbf{18} \\
 &&& Souffl\'e & 35 & 33 & 34 & 37 \\
 &&& Radlog & 254 & 295 & 340 & 164 \\
\midrule \midrule
\multirow{3}{*}{\textsc{ring10000}} & \multirow{3}{*}{10k} & \multirow{3}{*}{100,020,001} & \slog & 363 & 218 & 177 & \textbf{115} \\
 &&& Souffl\'e & 149 & 143 & 140 & 141 \\
 &&& Radlog & 464 & 646 & 852 & 1292 \\
\midrule \midrule
\multirow{3}{*}{\textsc{suitesparse}} & \multirow{3}{*}{412k} & \multirow{3}{*}{3,354,219,810} & \slog & -- & 1,593 & 908 & \textbf{671} \\
 &&& Souffl\'e & 1,417 & 1,349 & 1,306 & 1,282 \\
 &&& Radlog & -- & -- & -- & -- \\
\bottomrule
\end{tabular}
\label{tab:single-results}
\end{table}

Table~\ref{tab:single-results} details the results of our single-node
performance comparisons in seconds for each thread count, where each
datapoint represents the best of three runs (lower is
better). Experiments were cut off after 30 minutes. In every case, we
found that \slog{} produced the best performance overall at 120
threads, even compared to Souffl\'e's best time. However, as expected,
our results indicate that Souffl\'e outperforms \slog{} at lower core
counts (below 60). Souffl\'e implements joins with tight loops in
\CC{}, and (coupled with its superior single-node datastructures) this
allows Souffl\'e to achieve better performance than either \slog{} or
Radlog at lower core counts. We found that Radlog did not scale nearly
as well as either Souffl\'e or \slog{}. We expected this would be the
case: both \slog{} and Souffl\'e compile to \CC{}. By comparison,
Radlog's Spark-based architecture incurs significant sequential
overhead due to the fact that it is implemented on top of the JVM and
pays a per-iteration penalty by using Hadoop's aggregation and
shuffling phase. \slog{} also incurs sequential overhead compared to
Souffl\'e due to its distributing results after every iteration,
though results indicate that our MPI-based implementation helps
ameliorate this compared to Radlog.

\subsection{AAMs and CFAs}
\label{sec:eval:aam}
\label{sec:eval:pb2}

 \begin{table*}
 \centering
\caption{Control-Flow Analysis Experimental Results: Slog vs. Souffl\'e}\label{tab:results-aam}
 \begin{tabular}{ccccccccc}
   \toprule 
 & \multirow{2}{*}{Term Sz.}&\multirow{2}{*}{Iters}& \multirow{2}{*}{Cf. Pts}& \multirow{2}{*}{Sto. Sz.}& \multicolumn{2}{c}{$8$ Processes} & \multicolumn{2}{c}{$64$ Processes} \\
 & & & & & \multicolumn{1}{c|}{Slog} & \multicolumn{1}{c||}{Souffl\'e}& \multicolumn{1}{c|}{Slog}& \multicolumn{1}{c}{Souffl\'e}\\
 \midrule 
 \parbox[t]{2mm}{\multirow{6}{*}{\rotatebox[origin=c]{90}{3-$k$-CFA}}}
 & 8 & 1,193 & 98,114 & 23,413 & 00:01 & 01:07 & 0:02 & 00:15\\
 & 9 & 1,312 & 371,010 & 79,861 & 00:02 & 14:47 & 0:03 & 02:56 \\
 & 10 & 1,431 & 1,441,090 & 291,317 & 00:06 & \timeout{} & 0:05 & 45:49 \\
 & 11 & 1,550 & 5,678,402 &  1,107,957 & 00:27 & \timeout{} & 0:16 & \timeout{} \\
 & 12 & 1,669 & 22,541,634 & 4,315,125 & 02:14 & \timeout{} & 1:07 & \timeout{} \\
 & 13 & 1,788 & 89,822,530 & 17,022,965 & 12:17 & \timeout{} & 5:08 & \timeout{} \\
\midrule 
 \parbox[t]{2mm}{\multirow{6}{*}{\rotatebox[origin=c]{90}{4-$k$-CFA}}}
 & 9 & 1,363 & 311,790 & 65,397 & 00:01 & 14:38 & 00:03 & 02:08 \\
 & 10 & 1,482 & 1,197,038 & 229,621 & 00:05 & \timeout{} & 00:05 & 40:30 \\
 & 11 & 1,601 & 4,687,854 & 853,493 & 00:20 & \timeout{} & 00:13 & \timeout{} \\
 & 12 & 1,720 & 18,550,766 & 3,281,909 & 01:40 & \timeout{} & 00:53 & \timeout{} \\
 & 13 & 1,839 & 73,801,710 & 12,859,381 & 08:44 & \timeout{} & 03:58 & \timeout{} \\
 & 14 & 1,958 & 294,404,078 & 50,892,789 & 60:53 & \timeout{} & 35:46 & \timeout{} \\

\midrule 
 \parbox[t]{2mm}{\multirow{6}{*}{\rotatebox[origin=c]{90}{5-$k$-CFA}}}
 & 9 & 1,429 & 203,674 & 50,677 & 00:02 & 05:30 & 0:03 & 001:15 \\
 & 10 & 1,548 & 756,890 & 167,285 & 00:04 & 65:20 & 0:04 & 015:08 \\
 & 11 & 1,667 & 2,911,898  & 597,493 & 00:13 & \timeout{} &  0:08 & 196:06 \\
 & 12 & 1,786 & 11,416,218 & 2,245,109 & 00:56 &  \timeout{}& 0:27 & \timeout{} \\
 & 13 & 1,905 & 45,202,074 & 8,687,605 & 04:38 &  \timeout{}& 2:00 & \timeout{} \\
 & 14 & 2,024 & 179,882,650 & 34,158,581 & 25:14 & \timeout{}& 9:58 & \timeout{}\\
\midrule 
 \parbox[t]{2mm}{\multirow{6}{*}{\rotatebox[origin=c]{90}{10-$m$-CFA}}}
 & 50 & 6,120 & 21,005 & 656,847 & 00:02  & 00:02 & 00:10 & 00:01 \\
 & 100 & 11,670 & 42,855 & 2,781,447 & 00:07 & 00:09 & 00:20 & 00:04 \\
 & 200 & 22,770 & 86,555 & 11,440,647 & 00:26 & 00:56 &  00:42 & 00:23 \\
 & 400 & 44,970 & 173,955 & 46,399,047  & 01:44 & 06:26 & 01:38 & 01:56 \\
 & 800 & 89,370 & 348,755 & 186,875,847 & 07:35 & 45:22 &  04:21 & 09:33 \\
 & 1600 & 178,170 & 698,355 & 750,069,447 & 32:56 & \timeout{} & 14:36 & 62:35 \\
\midrule 
 \parbox[t]{2mm}{\multirow{6}{*}{\rotatebox[origin=c]{90}{12-$m$-CFA}}}
 & 25  & 3,559  & 17,105  & 385,279    & 00:01 & <0:01 & 0:06 & <0:01 \\
 & 50  & 6,434  & 36,280  & 1,885,179  & 00:04 &000:03 &  0:11 & 00:03 \\
 & 100 & 12,184 & 74,630  & 8,284,354  & 00:16 & 000:24 & 0:23 & 00:10 \\
 & 200 & 23,684 & 151,330 & 34,680,204 & 01:10 & 002:37 & 0:53 & 00:55 \\
 & 400 & 46,684 & 304,730 & 141,861,904 & 05:04 & 018:39 & 2:23 & 04:12 \\
 & 800 & 92,684 & 611,530  & 573,785,304 & 22:46 & 2:38:22 & 7:28 & 24:58 \\
\midrule 
 \parbox[t]{2mm}{\multirow{6}{*}{\rotatebox[origin=c]{90}{15-$m$-CFA}}}
 & 12  & 2,211  & 14,461  & 136,740     &  00:01 & <0:01 & 0:04 & <0:01 \\
 & 24  & 3,591  & 35,857  & 1,443,058   &  00:03 & 00:02 & 0:06 & 00:01 \\
 & 48  & 6,351  & 78,649  & 8,292,250   &  00:14 & 00:15 & 0:14 &  00:07 \\
 & 96  & 11,871 & 164,233 & 38,931,946  &  01:08 & 01:41 & 0:36 & 00:36 \\
 & 192 & 22,911 & 335,401 & 167,976,586 &  05:15 & 12:10 & 1:49 & 02:51 \\
 & 384 & 44,991 & 677,737 & 697,126,858 &  24:10 & 92:35 & 6:45 & 16:30 \\
 \bottomrule
 \end{tabular}
 \end{table*}

Next, we sought to benchmark the analyses described in
Section~\ref{sec:apps}, at scale, versus an equivalent
implementation using ADTs in Souffl\'e (we ignore Radlog in this
comparison due to its lack of support for ADTs). We developed a
\slog{} analysis for each of six different polyvariance choices: three
$k$-CFA ($k$=3,4,5) and three $m$-CFA ($m$=10,12,15)
implementations. We then systematically derived six different
Souffl\'e-based variants. We tested each of these on six different
term sizes, drawn from a family of worst-case terms identified in
David Van Horn's thesis~\cite{VanHorn:diss}. We then benchmark both \slog{} and Souffl\'e on each of these instances and report on their results,
scalability, and broad trends which we observed. Our
Souffl\'e code is an exact port of the \slog{} code we used (see Figure~\ref{fig:cesk-machine}), except
that \$-ADT values are used in place of subfacts and the analysis was
designed in the first place to avoid the need for these subfacts to
trigger rule evaluation as they can in \slog{}.

\paragraph*{Experimental Setup}

The experiments described in this subsection were run on a
\textsf{c6a.metal} instance rented from Amazon Web Services (AWS),
including 192 hardware threads (when run using the \textsf{.metal}
instance types) and 384 GiB of RAM. Because both \slog{} and Souffl\'e
are designed to enable parallelism, we ran each experiment at two
distinct scales: 8 and 64 processes (threads). \slog{} was invoked
using \textsf{mpirun}, and Souffl\'e's compiled mode was used to
produce a binary which was subsequently run and timed using GNU
\textsf{time}. We did not systematically measure memory usage; recent
microbenchmarks for TC report $3-5\times$ memory blowup versus Souffl\'e. We
record and report the best of three runs for each experiment (imposing
a four hour cutoff). To avoid an unfair comparison to Souffl\'e with
respect to on-disc ADT materialization (which may explode due to
linearization of linked data), our Souffl\'e implementation does not
output control-flow points or store directly---instead we measure and
report their size using the \textsf{sum} aggregate (built in to
Souffl\'e).

\paragraph*{Results}

We report our results in Table~\ref{tab:results-aam}. Each of six
distinct analysis choices is shown along the left side. Along rows of
the table, we show experiments for a specific combination of analysis,
precision, and term size. We detail the total number of iterations
taken by the \slog{} backend, along with control-flow points, store
size, and runtime at both $8$ and $64$ threads for \slog{} and
Souffl\'e. Times are reported in minutes / seconds form; several runs
of Souffl\'e took under 1 second (which we mark with $<$0:01), and
\timeout{} indicates that the run timed out after four hours.

Inspecting our results, we observed several broad trends. First, as
problem size increases, \slog{}'s runtime grows less-rapidly than
Souffl\'e's.
This point may be observed by inspecting runtimes for a specific
set of experiments. For example, 10-$m$-CFA with term size 200 took
\slog{} 26 seconds, while Souffl\'e's run took 56 seconds. Doubling
the term size to 400 takes 104 seconds in \slog{}, but 398 seconds in
Souffl\'e---a slowdown of $4\times$ in \slog{}, compared to a slowdown
of $7\times$ in Souffl\'e. A similar trend happens in many other
experiments, e.g., 15 minutes to over three hours for Souffl\'e
($13\times$ slowdown) vs. 2 to 4 seconds ($2\times$ slowdown) in
\slog{}'s runtime on 5-$k$-CFA. Inspecting the output of Souffl\'e's
compiled \CC{} code for each experiment helped us identify the source
of the slowdown. For example, the rule for returning a value to a continuation address \texttt{\$KAddr(e,env)}, in Figure~\ref{fig:kcfa-mcfa}, must
join a return state using this address with an entry in the continuation store for this address. Because Souffl\'e does not index subfacts, a scan-then-filter approach is used. In this case, as the subfact is an exact match, it could be optimized but in cases where a single variable in a subfact matches another, Souffle's scan-and-filter implementation cannot be avoided.

\begin{figure}
\small
\begin{Verbatim}[baselinestretch=.8]
// ret(av, k) :- ret(av, $KAddr(e, env)), kont_map($KAddr(e, env), k).
//        env0[0] ---^  env2[0]-^  ^--- env2[1]       env3[1] -----^      

if(!(rel_13_delta_ret->empty()) && !(rel_18_kont_map->empty())) {
  for(const auto& env0 : *rel_13_delta_ret) {
    RamDomain const ref = env0[1];
    if (ref == 0) continue;
    const RamDomain *env1 = recordTable.unpack(ref,2);
    {
      if( (ramBitCast<RamDomain>(env1[0]) == ramBitCast<RamDomain>(RamSigned(3)))) {
	RamDomain const ref = env1[1];
	if (ref == 0) continue;
	const RamDomain *env2 = recordTable.unpack(ref,2);
	{
	  for(const auto& env3 : *rel_18_kont_map) {
	    // On this line we've ommitted bitcasts and Tuple ctors:
	    if( !(rel_19_delta_kont_map->contains({{ {{ env2[0], env2[1] }}, env3[1] }}))
		&& !(rel_12_ret->contains({{env0[0], env3[1]}}))) {
	      // Omitted: null checks and insertion of {{ env0[0], env3[1] }} into ret
  }}}}}}}}}}
\end{Verbatim}
\caption{Example \CC{} code generated by Souffl\'e.}
\label{fig:souffle-cpp}
\end{figure}

This return rule and its compiled \CC{} code is shown
in Figure~\ref{fig:souffle-cpp}. Note that it uses two nested for loops to iterate over the entire \rtag{ref} relation, then iterate over the entire \rtag{kont\_map} relation, and finally check if there is a match for all possible combinations. In this way, Souffl\'e's lack of indices for structured values leaves it no other option but to materialize (in time) the entire Cartesian product as it checks for matches---rather than efficiently looking up relevant tuples in an appropriate index as would normally be the case for top-level relations.

For a fixed problem size, we found that Souffl\'e and \slog{} both
scaled fairly well. Souffl\'e consistently performed well on small
input sizes; additional processes did not incur slowdowns, and
Soufll\'e's efficiency was generally reasonable (roughly 50\%) when
algorithmic scalability did not incur slowdowns. For example, in
3-$k$-CFA (n=$8$), Souffl\'e took 67 seconds at 8 processes, and 15
seconds at 64 processes. \slog{}'s parallelism doesn't outweigh
communication overhead on smaller problems, particularly on problems
with high iteration count and low per-iteration work. As problem size
increases, our \slog{} implementations show healthy scalability;
efficiency grows as problem size grows (e.g., 24:10 to 6:45 on
15-$m$-CFA/384, 22:46 to 7:28 on 12-$m$-CFA/800).

We found that extracting optimal efficiency
from \slog{} was facilitated by increasing per-iteration work and
avoiding long sequences of sequential work with comparatively lower
throughput. Because each iteration represents a synchronization point, greater numbers of iterations will decrease opportunities for parallelism. As an example of this principle, our 10-$m$-CFA uses a
flat \rtag{ctx} fact to represent the context; a previous version used
a linked list $10$ elements deep, however this design achieved poorer
scaling efficiency due to these $10$-iteration-long sequences of work necessary to
extend the instrumentation at each call-site.
In our experiments scaling efficiency improved as polyvariance increased; e.g., improving by $2\times$ for 10-$m$-CFA, but $3.5\times$ for 15-$m$-CFA. We believe this is because of the relatively higher per-iteration work available with increased polyvariance.

\subsection{Multi-node Scaling Experiments}
\label{sec:eval:theta}
\label{sec:eval:pb3}
%

%% To test the ability of \slog{} to scale structured deductive inference
%% queries to high-performance compute workloads, we implemented an
%% analysis of $k$-CFA for the $\lambda$-calculus in \slog{} and
%% evaluated its scalability on Theta~\cite{van2008deciding}.  Our
%% implementation of $k$-CFA uses a relatively standard formulation
%% similar to that in~\cite{van2008deciding}, though our presentation
%% uses the $m$-top-stack-frames interpretation giving a
%% polynomially-increasing family of analyses for a fixed sensitivity
%% $m$~\cite{might2010resolving}. We evaluated our analysis using several
%% worst-case terms we generated to systematically grow the complexity of
%% the analysis to the highest-possible number of
%% states~\cite{van2008deciding}.

\begin{wrapfigure}{R}{2.7in}
\begin{center}
\vspace{-.05in}
\includegraphics[width=0.5\textwidth, trim={2cm 2cm 2cm
    2cm},clip]{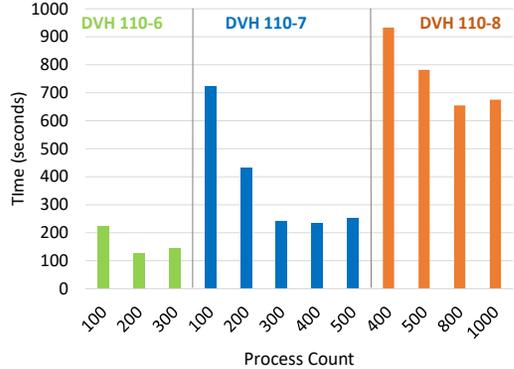}
\end{center}
\vspace{-.15in}
\caption{Strong-scaling experiments on Theta.}
\label{fig:theta}
\vspace{-.2in}
\end{wrapfigure}

In recent years, several cloud providers have launched MPI-capabable
HPC nodes for their cloud services and significantly upgraded their
network
interconnects~\cite{azure-august2019,parallelcluster}. Alongside
related advances in cloud architectures and MPI implementations, this
move signals the ability for massively parallel analytics to be
deployed as a service in the near future. In this spirit, we also
conduct some preliminary strong-scaling experiments on the Theta
supercomputer. For this we used a fully monomorphized $m$-CFA
(distinct from the $m$-CFA used in the previous
subsection)---experiments we had ready to go when our allocation on
Theta became possible. The Theta Supercomputer~\cite{parker2017early}
at the Argonne Leadership Computing Facility (ALCF) of the Argonne
National Laboratory is a Cray machine with a peak performance of
$11.69$ petaflops. It is based on the second-generation Intel Xeon Phi
processor and is made up of 281,088 compute cores. It has an aggregate
of $843.264$ TiB of DDR4 RAM, $70.272$ TiB of MCDRAM and $10$ PiB of
online disk storage. The supercomputer has a Dragonfly network
topology uses the Lustre filesystem.

We ran three sets of $m$-CFA worst-case experiments with $110$ terms
and $m$ = $6$, $7$ and $8$. The results of the three sets of
experiments, referred to as \texttt{dvh-110-6}, \texttt{dvh-110-7} and
\texttt{dvh-110-8} are plotted in Figure~\ref{fig:theta}. For all
three set of experiments we observe improvement in performance with
increase in process counts, until maximum efficiency is attained,
after which performance degrades with increasing process counts, due
to communication overhead and workload starvation. In general, for a
given workload (problem size), we observe a range of processes that
exhibit healthy scalability. \texttt{dvh-110-6} shows a near $100\%$
scaling efficiency ($2\times$ speedup while increasing the process
count from $100$ to $200$), performance however drops when the number
of processes is increased to 300. Similarly, \texttt{dvh-110-7} shows
a $75\%$ scaling efficiency ($3\times$ speedup when the process count
is increased from $100$ to $400$), and \texttt{dvh-110-8} shows a
$71\%$ scaling efficiency ($1.42\times$ speedup going $400$ to $800$).

%Even though the largest problem is run at $2\times$ time process count,
%the scaling efficiency drops by only $4\%$, indicative of a robust and a
%scalable parallel deductive inference backend.\tom{last sentence is awful...}

\section{Related and Future Work}
\label{sec:related}

\paragraph*{Distributed Datalog}

There have been significant implementation efforts to scale
Datalog-like languages to large clusters of machines. For example,
BigDatalog~\cite{bigdl}, Distributed SociaLite~\cite{Seo:2013},
Myria~\cite{Halperin:2014}, and Radlog~\cite{Gu:2019} all run on
Apache Spark clusters (servers networked together via commodity
switches within a datacenter). Extending Spark's architecture with
recursive queries (and aggregates), these frameworks scale to large
datasets typical of Spark queries. \slog{} differs from these systems
in two primary ways. First, compared to \slog{}'s MPI-based
implementation, Apache Spark's framework-imposed overhead is
increasingly understood to be a bottleneck in scalable data analytics
applications, with several authors noting order-of-magnitude
improvements when switching from Spark to
MPI~\cite{SparkBadMPIGood,Kumar:2017,Anderson:2017}. Second,
none of the aforementioned systems support first-class subfacts; for
example, while Radlog can compute the length of the shortest path from
a specific point, it cannot materialize the path per-se. Recently,
Radlog's authors have created DCDatalog, a parallel Datalog which
targets shared-memory SMP architectures and demonstrate a $10\times$
runtime speedup compared to Souffl\'e on a machine with four
eight-core processors and 256GB of RAM. Unfortunately, DCDatalog is
not open-source, and we have not been able to obtain a copy for
evaluation; we believe it is difficult to interpret DCDatalog's
results compared to \slog{} and Souffl\'e, as their paper notes
``Souffl\'e does not allow aggregates in recursion, and thus it must
use a stratified query that results in very poor performance'' for
several evaluation queries. Last, Nexus (also closed-source) has
claimed a significant performance boost (up to $4\times$) compared to
BigDatalog by using Apache Flink, a data-flow processing
language~\cite{Muhammad:2022}.

\paragraph*{Datalog Extensions}

Noting the first-order nature of vanilla Datalog---and often inspired
by Datalog's efficient semi-na\"ive evaluation strategy---there has
been extensive work in extending Datalog with additional expressive
power~\cite{flix-madsen2016,Madsen:2018,formulog-bembenek2020,Arntzenius:2016,Arntzenius:2019}. Flix
augments Datalog with lattices~\cite{flix-madsen2016,Madsen:2018}, but
is not specifically focused on efficient compilation; recently, Ascent
is a macro-based implementation of Datalog in Rust which includes
lattices and shows orders-of-magnitude runtime improvements versus
Flix~\cite{Sahebolamri:2022}. Similarly, Datafun is a pure functional
language which computes fixed points of monotone maps on
semilattices~\cite{Arntzenius:2016,Arntzenius:2019}. Compared to
\slog{}, Datafun's evaluation strategy is top-down and based on the
$\lambda$-calculus; the authors have recently studied semi-na\:ive
evaluation of Datafun upon work on the incremental
$\lambda$-calculus~\cite{Giarrusso:2019,Yufei:2014}. \slog{}'s primary
difference from this work is that it is based on \core{} rather than
the $\lambda$-calculus; because of this, semi-na\"ive evaluation for
functions in \slog{} (using defunctionalization) requires no extra
logic.

\paragraph*{Datalog + Constraints}

An increasingly-popular semantic extension to Datalog is adding
first-class
constraints~\cite{formulog-bembenek2020,Madsen:2020,Emina:2013,Torlak:2014}. These
constraints typically allow interfacing with an SMT solver,
potentially in a loop with subsequent
analysis~\cite{formulog-bembenek2020}.  Formulog includes ADTs and
first-order functions over ADTs, allowing Turing-equivalent to build
formulas of arbitrary size to be checked by
Z3~\cite{formulog-bembenek2020}; we anticipate \slog{} will perform
well compared to Formulog when subfact-indexing is of concern, though
by Amdahl's law this effect will be smaller in code whose computation
is dominated by calls to Z3. Similarly, Rosette efficiently compiles
solver-aided queries to efficient implementations using host language
constructs and a symbolic virtual machine
(SVM)~\cite{Emina:2013,Torlak:2014}. \slog{} is largely orthogonal to
these systems, which focus on shared-memory implementations and are
not primarily concerned with parallelization. We have transliterated
proof-of-concept examples from both of these projects into \slog{},
but it is currently impossible to call Z3 from \slog{} as doing so
would require all facts be resident on a single node. Semantically,
\slog{} is more directly comparable to constrained HornSAT or
existential fixed-point logic, which have attracted recent interest
for their application to program
verification~\cite{Fedyukovich:2018,Blass1987,Bjorner:2015,Arie:2022}. \core{}
can express constrained HornSAT problems as long as a decision
procedure for the background logic is available; we plan to study
usage of \slog{} for CHCs in subsequent work.

%% \paragraph*{Implementation Languages for Formal Systems}

%% There have been many efforts to design languages that allow
%% expressing, reasoning about, and implementing formal
%% systems~\cite{,ROSU2010397,Matthias:2009,

%% \paragraph*{Equality Saturation and Congruence Closure}

%% Like our implementation of \lf{}, many other formal systems rely upon
%% congruence closure. Tate et al. introduced the concept of equality
%% saturation, which simultaneously applies rewriting to each relevant
%% term~\cite{Tate:2009}. Equality saturation was further improved and
%% implemented in the modern Rust library Egg~\cite{Willsey:2021}. We
%% believe \slog{} will be an ideal fit for congruence closure, because
%% we believe applying rewriting will parallelize very well given
%% \slog{}'s design. Recent success in SMT solving also relies upon
%% equality saturation, and we believe successful parallelization of
%% equality saturation will represent a first step towards
%% massively-parallel SMT solving~\cite{smtz3}.

\paragraph*{Parallel Program Analyses}

Given the algorithmic complexity intrinstic to large-scale program
analyses, there has been significant interest in its
parallelization~\cite{Stefan:2011,Aiken:2007,Xie:2007,Siddiqui:2010,antoniadis2017porting,Bravenboer:2009}
or implementation using special-purpose
datastructures~\cite{Whaley:2005,Prabhu:2011,Kramer:1994,Lee:90,MendezLojo:2010}. There
are a variety of fundamental approaches to scalability; for example,
summarization-based analyses (such as
Saturn~\cite{Aiken:2007,Xie:2007}) are attractive due to the
task-level parallelism they expose. Much work in scaling program
analysis has focused on context-insensitive analyses---wherein
task-level parallelism is more directly exploitable. The goals of
\slog{} are most closely related to current efforts on scaling rich,
whole-program context-sensitive analyses using deductive
inference~\cite{antoniadis2017porting,10.1007/978-3-319-41540-6,Scholz:2016:FLP:2892208.2892226}.

\paragraph*{Parallel Relational Algebra}

\slog{}'s backend builds upon recent successes in balanced, parallel
relational algebra (BPRA) and follow-up work on compilation of vanilla
Datalog to parallel relational algebra
kernels~\cite{loadbalancingra,hipc,cc}. However, that work focuses
mainly on the low-level implementation of relational algebra kernels
rather than a unified programming language, compiler, and runtime.

\section{Conclusion}
\label{sec:conclusion}

%Datalog is an increasingly-popular implementation language for static
%analyses and other formal systems due to its balance of expressivity
%and performance.
In this work we extended Datalog with subfacts,
explicating a core language, \core{}, which supports subfacts as
first-class facts, values, and threads. This straightforward semantic extension enables
several key innovations, including the expression of higher-order
functions via defunctionalization, subfact indexing,
and a rich connection to constructive logic via the proofs-as-programs
interpretation. We have implemented these ideas in \slog{}, a fully featured language
for data-parallel structured deduction, using a new MPI-based relational algebra backend,
and we demonstrate its application to operational semantics, abstract interpreters, and
formal systems broadly. Our experiments show that \slog{} is competitive with,
or (when programming with structured heap data) orders-of-magnitude faster
than state-of-the-art systems, showing improved data-parallelism and scaling
up to hundreds of cores on large unified machines and high-performance clusters.

\bibliographystyle{plain}
\bibliography{local}

\end{document}
%\endinput
%%
%% End of file `sample-acmsmall-conf.tex'.